\documentclass[letterpaper]{article} 
\usepackage{aaai2026}  
\nocopyright  
\usepackage{times}  
\usepackage{helvet}  
\usepackage{courier}  
\usepackage[hyphens]{url}  
\usepackage{graphicx} 
\usepackage{natbib}  
\usepackage{caption} 
\usepackage{algorithm}

\usepackage{newfloat}
\usepackage{listings}

\usepackage{graphicx}
\usepackage{amsmath}
\usepackage{amsmath,amssymb,amsfonts}
\usepackage{amssymb}
\usepackage{breqn}
\usepackage{ragged2e}
\usepackage{cleveref}
\usepackage{cases}
\usepackage{graphicx}
\usepackage{multirow}
\usepackage{mathrsfs}
\usepackage{textcomp}
\usepackage{xcolor}
\usepackage{subfigure}
\usepackage{bm}
\usepackage{multicol} 
\usepackage{caption}
\usepackage{algorithm}
\usepackage{algpseudocode}
\usepackage{amsthm}
\usepackage{booktabs} 
\usepackage{multirow} 
\usepackage{graphicx} 
\usepackage{makecell}

\DeclareCaptionStyle{ruled}{labelfont=normalfont,labelsep=colon,strut=off} 
\floatstyle{ruled}
\newfloat{listing}{tb}{lst}{}
\floatname{listing}{Listing}
\title{Fading the Digital Ink: A Universal Black-Box Attack Framework for 3DGS Watermarking Systems}
\author {
    Qingyuan Zeng\textsuperscript{\rm 1},
    Shu Jiang\textsuperscript{\rm 1},
    Jiajing Lin\textsuperscript{\rm 2},
    Zhenzhong Wang\textsuperscript{\rm 2},
    Kay Chen Tan\textsuperscript{\rm 3},
    Min Jiang\textsuperscript{\rm 2}\thanks{Corresponding author.}
}
\affiliations {
    \textsuperscript{\rm 1} Institute of Artificial Intelligence, Xiamen University, China\\
    \textsuperscript{\rm 2} School of Informatics, Xiamen University, China\\
    \textsuperscript{\rm 3} Department of Data Science and Artificial Intelligence, The Hong Kong Polytechnic University, Hong Kong SAR\\
}

\usepackage{bibentry}

\begin{document}

\maketitle

\begin{figure*}
    \centering
    \includegraphics[width=1\linewidth]{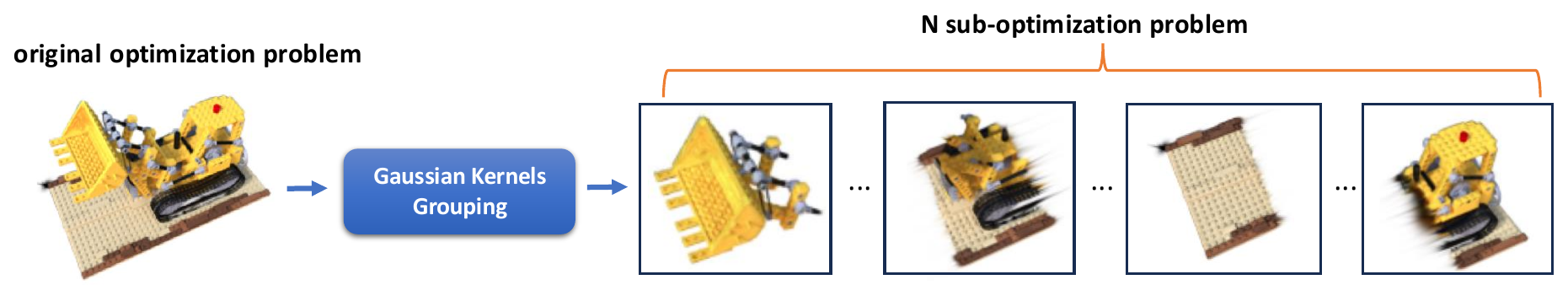}
    \caption{Illustration of the group-based optimization strategy. The original 3DGS model (full Lego object) is partitioned into multiple sub-optimization problems (individual Lego components) to manage the large search space effectively.}
    \label{fig:grouping}
\end{figure*}

\begin{abstract}

With the rise of 3D Gaussian Splatting (3DGS), a variety of digital watermarking techniques, embedding either 1D bitstreams or 2D images, are used for copyright protection. However, the robustness of these watermarking techniques against potential attacks remains underexplored. This paper introduces the first universal black-box attack framework, the Group-based Multi-objective Evolutionary Attack (GMEA), designed to challenge these watermarking systems. We formulate the attack as a large-scale multi-objective optimization problem, balancing watermark removal with visual quality. In a black-box setting, we introduce an  indirect objective function that blinds the watermark detector by minimizing the standard deviation of features extracted by a convolutional network, thus rendering the feature maps uninformative. To manage the vast search space of 3DGS models, we employ a group-based optimization strategy  to partition the model into multiple, independent sub-optimization  problems. Experiments demonstrate that our framework effectively removes both 1D and 2D watermarks from mainstream 3DGS watermarking methods while maintaining high visual fidelity. This work reveals critical vulnerabilities in existing 3DGS copyright protection schemes and calls for the development of more robust watermarking systems.

\end{abstract}


\section{Introduction}

3D Gaussian Splatting (3DGS) is an emerging 3D scene representation and reconstruction technology. It has the advantages of high fidelity, fast rendering speed, and real-time rendering capabilities \cite{kerbl3Dgaussians,10897713}. It has broad application prospects in fields such as film production, game development, virtual reality, and autonomous driving \cite{10.5555/3737916.3741145,Tu2025VRSplat,chen2025omnire}. Given that creating a 3DGS model represents a significant investment in data acquisition, engineering expertise, and computational resources \cite{zhang2024physdreamer}, and the copyright of 3DGS assets is prone to unauthorized distribution and malicious tampering, it is crucial to effectively protect the copyright of 3DGS assets. To meet this challenge, various invisible watermarking methods for 3DGS have been proposed \cite{chen2025guardsplat,jang20253dgsw3dgaussiansplatting,huang2024gaussianmarker}. They embed invisible copyright information directly into the Gaussian parameters of 3DGS models, which is subsequently extracted from the rendered images. These approaches encompass different strategies, such as encoding one-dimensional (1D) copyright strings \cite{chen2025guardsplat} or hiding entire two-dimensional (2D) data like logos or images as watermarks \cite{NEURIPS2024_59091e82}.

These 3DGS invisible watermarking methods \cite{chen2025guardsplat,huang2024gaussianmarker}  need to be robust enough to truly protect the copyright of 3D assets in reality. That is, they need to maintain the integrity and detectability of watermarks in the face of various potential attacks \cite{cox1997secure,zhao2024invisible}.  However, so far, no research has explored the robustness of 3DGS invisible watermarks when facing attacks. Therefore, this paper aims to answer the following question: is there a universal attack method that can destroy these 3DGS invisible watermarks?

There are the following difficulties in destroying 3DGS invisible watermarks. First, visual fidelity must be preserved. The attacker is required to remove the watermark without noticeably degrading the quality of the 3DGS model \cite{zhang2020survey}. Second, attacks typically occur in a black-box setting, where attackers often lack knowledge of the watermark's content, embedding process, and detection process \cite{papernot2017practical,zeng2024ask}. This means an attacker cannot accurately locate the watermark's distribution in the Gaussian parameters or rendered images, nor can they use the watermark detector's gradients to guide the optimization process through backpropagation.

To narrow this research gap, we propose GMEA, a universal black-box attack framework against 3DGS invisible watermarks. Our approach formulates the attack as a large-scale multi-objective optimization problem \cite{wang2024generating}, seeking to simultaneously destroy the watermark while preserving the model's visual quality \cite{wang2021survey}. The decision variables represent the attacker's two primary actions: selectively pruning some of the model's Gaussian kernels and subtly shifting the color values of others. The optimization objectives are: 1) minimizing visual quality degradation measured by MSE between original and perturbed renders, and 2) maximizing watermark destruction. Critically, in the black-box setting where watermark detectors are inaccessible, we introduce an indirect objective function to evaluate watermark destruction: the standard deviation of convolutional feature maps extracted from rendered images \cite{lu2020enhancing}. Minimizing this deviation reduces discriminative information in feature maps, effectively blinding downstream watermark decoders that rely on convolutional features to extract watermark signals. This breaks the watermark extraction process despite having no detector access.

\begin{figure*}
    \centering
    \includegraphics[width=0.95\linewidth]{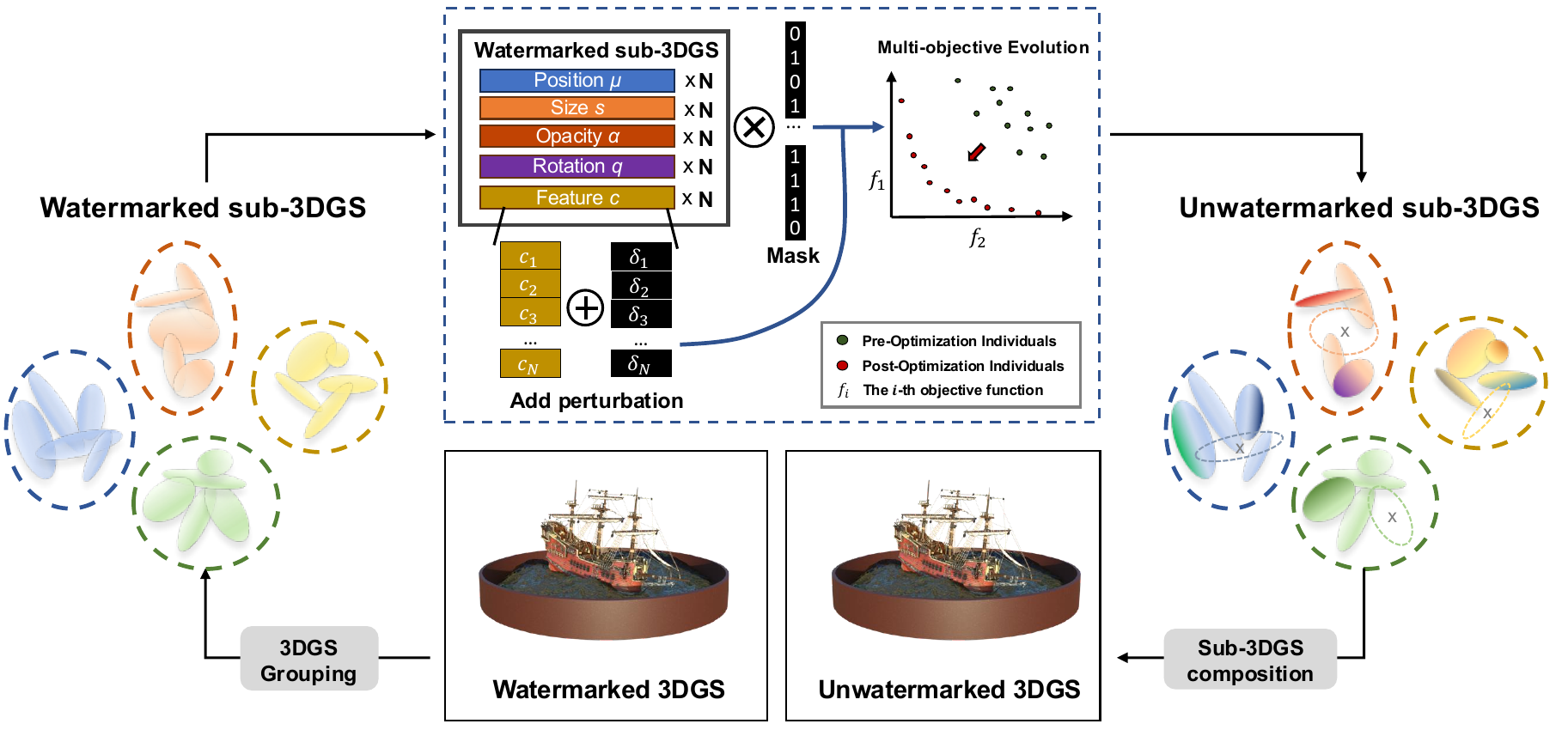}
    \caption{An overview of our proposed Group-based Multi-objective Evolutionary Attack (GMEA) framework. The attack pipeline begins by partitioning the watermarked 3DGS model into several spatially coherent sub-3DGS. Each sub-3DGS then undergoes an independent multi-objective evolutionary optimization. During this phase, potential modifications—represented by a binary mask for pruning Gaussians and a perturbation vector for altering colors—are evolved to simultaneously minimize visual quality loss ($F_1$) and maximize watermark destruction ($F_2$). The resulting optimized and unwatermarked sub-3DGS are then reassembled to form the final unwatermarked 3DGS model.}
    \label{fig:pipeline}
\end{figure*}

To solve this large-scale multi-objective problem, we build GMEA upon an evolutionary algorithm, inspired by its unique ability to handle such complex optimizations \cite{deb2002fast,hong2023improving,9723472,wang2024generating}. However, a direct application of evolutionary algorithms is computationally inefficient. Due to the oversized search space created by the vast number of Gaussian kernels, the process of converging to an effective solution is slow.  Therefore, to make the optimization tractable and efficient, we break down the large-scale optimization problem into several smaller sub-optimization problems to be solved independently \cite{7544478}, as shown in Figure \ref{fig:grouping}. Then, we combine the solutions of each sub-optimization problem to obtain the solution to the original optimization problem. Specifically, as shown in Figure \ref{fig:pipeline}, we use unsupervised clustering algorithms (such as K-Means \cite{MacQueen1967SomeMF,10.1007/978-3-030-58607-2_16}) to cluster the Gaussian kernels of the watermarked 3DGS model into k clusters from the perspective of position, obtaining k sub-3DGS models. Then, we perform multi-objective evolutionary algorithm on each sub-3DGS model to remove the watermark. Finally, we merge all the optimized sub-3DGS models to obtain the complete 3DGS model without watermarks. Our contributions can be summarized as follows:

\begin{enumerate}

    \item We propose the first universal black-box attack framework GMEA against 3DGS invisible watermarks. It features a model-agnostic objective that disables watermark detection by disrupting convolutional features without requiring any knowledge of the detector.
    
    \item We design a group-based optimization strategy that partitions the immense search space of 3DGS models, significantly improving the search efficiency of our evolutionary algorithm in discovering effective solutions.
    \item We conduct extensive experiments to validate our framework's effectiveness and universality, successfully attacking leading methods for both 1D and 2D 3DGS watermarking.

\end{enumerate}

\section{Related Works}

\subsection{3D Gaussian Splatting}

3D Gaussian Splatting (3DGS) is a foundational technique for real-time, photo-realistic rendering, with its applications rapidly expanding \cite{kerbl3Dgaussians,zhu20243dgaussiansplattingrobotics,10870258}. In 3D content generation, frameworks like GaussianDreamer utilize 2D diffusion priors for automated asset creation from text or images \cite{10655860}. For dynamic scenes, 4D Gaussian Splatting models complex, non-rigid motions via time-varying deformation fields \cite{Wu_2024_CVPR}. Its utility extends to geometric reconstruction, where methods like SuGaR extract detailed meshes by imposing local geometric constraints \cite{guedon2023sugar}. Furthermore, PhysGaussian enables realistic physical interactions by treating Gaussians as Lagrangian particles within a physics simulator \cite{xie2023physgaussian}.


As 3DGS integrates with physics engines and generative models, becoming more complex and less transparent, there's an urgent need in academia and industry to establish copyright protection, robustness analysis, and trust-validation systems for these models.

\subsection{3D Gaussian Splatting Watermarking}
Modern 3DGS watermarking techniques imperceptibly embed copyright data while preserving visual fidelity. Approaches are typically categorized by message dimensionality: 1D bitstreams for simple copyright strings, and 2D messages for complex data like logos or images.

\subsubsection{1D Bitstream Watermarking.}

These methods focus on embedding one-dimensional (1D) binary messages, such as copyright strings, directly into the 3DGS model. GuardSplat \cite{chen2025guardsplat} embeds messages by modifying existing Gaussian kernels. It utilizes small, learnable offsets to the Spherical Harmonic (SH) features to integrate the watermark, a technique designed to preserve the model's original 3D structure and visual fidelity. 3D-GSW \cite{jang20253dgsw3dgaussiansplatting} takes a refinement approach, preparing the model for watermarking through a process called Frequency-Guided Densification (FGD). This technique first prunes Gaussians that have minimal impact on rendering quality and then splits Gaussians located in high-frequency areas to maintain visual fidelity. The watermark is subsequently embedded into this optimized set of Gaussians via fine-tuning. GaussianMarker \cite{huang2024gaussianmarker} employs an additive strategy, leaving original Gaussians untouched. It identifies regions of high uncertainty in the model and introduces new, dedicated Gaussian kernels called 'GaussianMarkers' within these areas to carry the watermark information.

\subsubsection{2D Message Watermarking.}

Another approach pushes the capacity of steganography further by enabling the embedding of two-dimensional (2D) messages, such as entire images or logos. GS-Hider \cite{NEURIPS2024_59091e82} exemplifies this category. It achieves high-capacity message hiding by fundamentally altering the rendering pipeline. Instead of just modifying SH coefficients, it replaces them entirely with a coupled secured feature attribute. This high-dimensional feature is then rendered into a feature map. Two parallel decoders are employed: a public scene decoder that reconstructs the original visual scene, and a private message decoder that extracts the hidden 2D image from the same feature map. This decoupling allows for the concealment of complex messages without direct interference with the primary scene's rendering. 

While the practicality of watermarking methods depends on robustness, the resilience of 3DGS schemes against dedicated attacks remains largely untested. We therefore introduce the first universal attack framework to serve as a security benchmark for the 3DGS ecosystem. Our work aims to not only assess current vulnerabilities but also to catalyze the development of more secure future solutions.

\section{Methodology}

In this section, we detail our proposed black-box attack framework, Group-based Multi-objective Evolutionary Attack (GMEA), designed to remove invisible watermarks from 3DGS models. The overall pipeline of our method is illustrated in Figure \ref{fig:pipeline}, while the complete pseudocode and theoretical proofs \underline{are provided in the Appendix.}

\subsection{Problem Formulation}

We aim to find an adversarial 3DGS model, $\mathbf{G}_{adv}$, derived from a watermarked 3DGS model $\mathbf{G}_{wm}$. The goal is to simultaneously maintain high visual fidelity and destroy the embedded watermark. This is formulated as a multi-objective optimization problem:

\begin{equation}
\begin{aligned}
\min_{\mathbf{G}_{adv}} \quad & { F_1(\mathbf{G}_{adv}, \mathbf{G}_{wm}), F_2(\mathbf{G}_{adv}) } \\
\text{s.t.} \quad & \mathbf{G}_{adv} \in \mathcal{P}(\mathbf{G}_{wm})
\end{aligned}
\end{equation}

Here, $F_1$ measures visual quality loss and $F_2$ quantifies watermark destruction. The constraint $\mathbf{G}_{adv} \in \mathcal{P}(\mathbf{G}_{wm})$ specifies that any candidate solution $\mathbf{G}_{adv}$ must be generated by applying the allowed perturbations to $\mathbf{G}_{wm}$.

\subsection{Group-Based Optimization Strategy}

Our group-based optimization strategy tackles the immense search space of 3DGS models    by decomposing the task into smaller, spatially coherent sub-problems. This partitioning makes the search for an effective solution more efficient by simplifying the optimization landscape.

To achieve this partitioning, we employ the K-Means clustering on the spatial coordinates ($\mathbf{xyz}$) of the Gaussian kernels. Let the set of all 3D coordinates for the $N$ Gaussians in the watermarked model $\mathbf{G}_{wm}$ be $\mathbf{P} = \{\mathbf{p}_1, \mathbf{p}_2, \dots, \mathbf{p}_N\}$, where $\mathbf{p}_j \in \mathbb{R}^3$. The goal of K-Means is to partition this set of kernels $\mathbf{P}$ into $k$ disjoint spatial clusters, denoted by $\mathbf{S} = \{S_1, S_2, \dots, S_k\}$, by minimizing the within-cluster sum of squares. The minimization criterion is formalized as:
\begin{equation}
\underset{\mathbf{S}}{\arg\min} \sum_{i=1}^{k} \sum_{\mathbf{p}_j \in S_i} \|\mathbf{p}_j - \boldsymbol{\mu}_i\|^2,
\end{equation}
where $\boldsymbol{\mu}_i$ is the geometric centroid of the coordinates in cluster $S_i$.

Once the optimal coordinate clusters $\{S_1, \dots, S_k\}$ are determined, we map these spatial groupings back to the Gaussians' original indices. This creates a corresponding partition of the index set $\{1, 2, \dots, N\}$ into $k$ disjoint sets $\{I_1, \dots, I_k\}$. Each index set $I_i$ is formalized as
\begin{equation}
I_i = \{j \mid \mathbf{p}_j \in S_i\}.
\end{equation}

With the indices properly partitioned, we formally construct the 3DGS sub-models. The $i$-th sub-model, $\mathbf{G}^{(i)}_{wm}$, is defined as the collection of Gaussian kernels whose indices fall into the set $I_i$:
\begin{equation}
\mathbf{G}^{(i)}_{wm} = \{g_j \in \mathbf{G}_{wm} \mid j \in I_i \},
\end{equation}
where $g_j$ represents the $j$-th Gaussian kernel. This decomposition of the large-scale task into smaller sub-problems significantly reduces search complexity, enabling a more efficient optimization.

\subsection{Multi-objective Evolutionary Attack}

Each sub-problem defined by a sub-model $\mathbf{G}^{(i)}_{wm}$ is solved with a multi-objective evolutionary algorithm to find an effective adversarial solution , \(\mathbf{G}^{(i)}_{adv}\). The algorithm refines a population of solutions to approximate the Pareto-optimal front, balancing the conflicting objectives of visual fidelity and watermark removal.

\subsubsection{Individual Representation.}

Each individual in our population represents a potential modification to a sub-model $\mathbf{G}^{(i)}_{wm}$ containing $N^{(i)}$ Gaussians. The individual's genetic representation is a vector
\begin{equation}
\mathbf{x}^{(i)} = [\mathbf{m}^{(i)}, \mathbf{c}^{(i)}],
\end{equation}
which concatenates a binary mask vector $\mathbf{m}^{(i)} \in \{0, 1\}^{N^{(i)}}$ and a color perturbation vector $\mathbf{c}^{(i)} \in [-\epsilon, \epsilon]^{3N^{(i)}}$. The mask vector determines which Gaussian kernels are pruned ($m_{j}=0$), while the color perturbation vector defines additive shifts to the DC color component (RGB) for the remaining Gaussian kernels.

The adversarial sub-model $\mathbf{G}^{(i)}_{adv}$ is constructed by applying these modifications. Let $\mathbf{C}_{dc}$ be the $N^{(i)} \times 3$ matrix of original DC colors. The new color matrix, $\mathbf{C}_{dc, adv}$, is computed as:
\begin{equation}
\mathbf{C}_{dc, adv} = \text{diag}(\mathbf{m}^{(i)}) \left( \mathbf{C}_{dc} + \text{Reshape}(\mathbf{c}^{(i)}) \right).
\end{equation}

Here, the diagonalized mask $\text{diag}(\mathbf{m}^{(i)})$ filters the Gaussian kernels, while the reshaped perturbation vector modifies the colors of the survivors. All other Gaussian parameters (e.g., position, scale, opacity) are similarly filtered by the mask. The resulting adversarial sub-model $\mathbf{G}^{(i)}_{adv}$ is then used to render images $\mathbf{R}_{adv}^{(i)}$ for fitness evaluation.






\begin{figure*}
    \centering
    \includegraphics[width=1\linewidth]{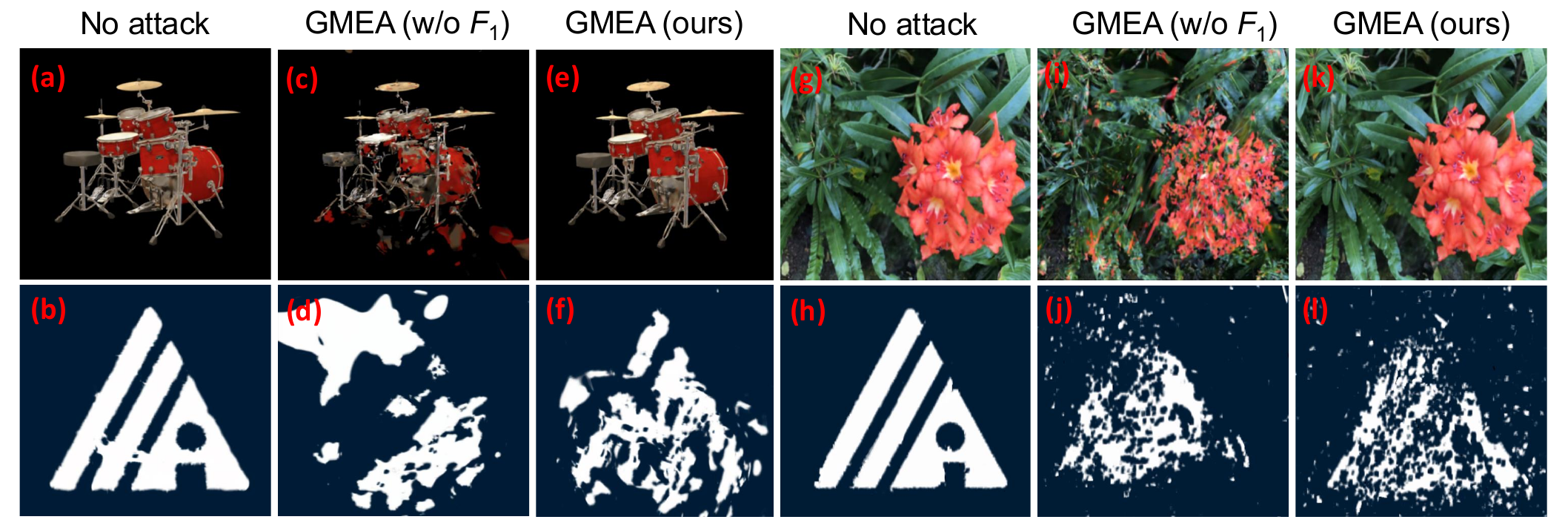}
    \caption{Qualitative results of the GMEA attack on the \textit{Drums} (a-f) and \textit{Flower} (g-l) scenes. While a single-objective attack GMEA (w/o $F_1$) corrupts the watermark at the cost of severe visual degradation (c,d,i,j), our full multi-objective approach GMEA successfully removes the watermark while preserving high visual fidelity (e,f,k,l).}
\label{fig:qualitative_comparison}
\end{figure*}

\subsubsection{Objective Functions.}

We evaluate each individual's fitness using two conflicting objectives designed to preserve visual fidelity while removing the watermark. 

The first objective, visual quality loss ($F_1$), measures the perceptual difference between the original watermarked 3DGS and adversarial 3DGS. We calculate this loss as a weighted combination of the L1 distance and the Structural Similarity Index Measure (SSIM) over images rendered from multiple viewpoints $\{v_1, \dots, v_{N_v}\}$:

\begin{equation}
\begin{split}
F_1(\mathbf{G}_{adv}^{(i)}) = \frac{1}{N_v} \sum_{v=1}^{N_v} \Big[ \lambda \mathcal{L}_{\text{L1}}(\mathbf{R}_{adv}^{(i,v)}, \mathbf{R}_{wm}^{(i,v)}) \\
+ (1-\lambda) (1 - \text{SSIM}(\mathbf{R}_{adv}^{(i,v)}, \mathbf{R}_{wm}^{(i,v)})) \Big],
\end{split}
\end{equation}
where $\mathbf{R}_{adv}^{(i,v)}$ and $\mathbf{R}_{wm}^{(i,v)}$ are the rendered images of adversarial 3DGS and original watermarked 3DGS.  $\lambda$ is a weighting factor. Minimizing $F_1$ guides solutions to be visually indistinguishable from the original.

The second objective, watermark destruction ($F_2$), provides a model-agnostic attack by neutralizing the convolutional feature extraction common to all decoders. We achieve this by minimizing the feature maps' statistical variance, which flattens their patterns and renders them non-discriminative for watermark detection. Specifically, we pass a rendered adversarial image $\mathbf{R}_{adv}^{(i,v)}$ through a convolutional feature extractor $\Phi$. The dispersion of a single feature channel, $D(\mathbf{F}_c)$, is then quantified by its standard deviation:
\begin{equation}
D(\mathbf{F}_c) = \sqrt{\frac{1}{H'W'}\sum_{h,w}(\mathbf{F}_c(h,w) - \bar{\mathbf{F}}_c)^2},
\end{equation}
where $\bar{\mathbf{F}}_c$ is the mean activation of channel $c$. The final objective $F_2$ is the average dispersion over all feature channels and viewpoints:
\begin{equation}
F_2(\mathbf{G}_{adv}^{(i)}) = \frac{1}{N_v} \sum_{v=1}^{N_v} \left[ \frac{1}{C'} \sum_{c=1}^{C'} D(\Phi(\mathbf{R}_{adv}^{(i,v)})_c) \right].
\end{equation}


    
        
        
    
    


\subsubsection{Evolutionary Process.}

The evolutionary process begins with a population $\mathcal{P}_t$ of size $N_{pop}$. At each generation $t$, an offspring population $\mathcal{Q}_t$ is generated from the current population $\mathcal{P}_t$. This involves two main operators. First, a crossover operator produces two new solutions from a pair of parents ($\mathbf{x}_a, \mathbf{x}_b$), distributing the offspring around the parents' positions in the search space:
\begin{equation}
\mathbf{x}'_{a,b} = 0.5 \left[ (\mathbf{x}_a+\mathbf{x}_b) \mp \beta|\mathbf{x}_b-\mathbf{x}_a| \right],
\end{equation}
where $\beta$ is a hyperparameter controlling the spread of the offspring. Subsequently, a mutation operator introduces fine-grained perturbations to an individual $\mathbf{x}'$ to enhance local exploration:
\begin{equation}
x''_j = x'_j + \eta_j(ub_j - lb_j),
\end{equation}
where $\eta_j$ is a small perturbation value, and $[lb_j, ub_j]$ represents the defined lower and upper bounds for the $j$-th decision variable.

After generating offspring, a rigorous selection process determines which individuals will form the next generation's population. First, the current parent population ($\mathcal{P}_t$) and their offspring ($\mathcal{Q}_t$) are merged into a combined pool, $\mathcal{R}_t = \mathcal{P}_t \cup \mathcal{Q}_t$. This pool is then ranked and partitioned into a hierarchy of non-dominated fronts $\{\mathcal{F}_1, \mathcal{F}_2, \dots\}$ based on Pareto dominance \cite{hong2023improving}.

The next generation is formed by elitism, admitting individuals from the best non-dominated fronts ($\mathcal{F}_1, \mathcal{F}_2, \dots$) until the population capacity ($N_{pop}$) is met. To maintain diversity when the final front ($\mathcal{F}_l$) is truncated, we rank its members by a density metric that prioritizes solutions in less crowded regions. The density score for an individual solution $\mathbf{x}$, denoted $d(\mathbf{x})$, is calculated as:
\begin{equation}
d(\mathbf{x}) = \sum_{o=1}^{M} \frac{F_o(\text{neighbor}^+) - F_o(\text{neighbor}^-)}{F_o^{\max} - F_o^{\min}},
\end{equation}
where $M$ is the number of objectives, and $F_o(\text{neighbor}^{\pm})$ are the objective values of the neighbors of solution $\mathbf{x}$ after sorting the front along objective $o$. Individuals with higher density scores are chosen to fill the remaining slots, forming a diverse parent population for the next evolutionary cycle.

\begin{figure*}[htbp!]
    \centering
    \includegraphics[width=1\linewidth]{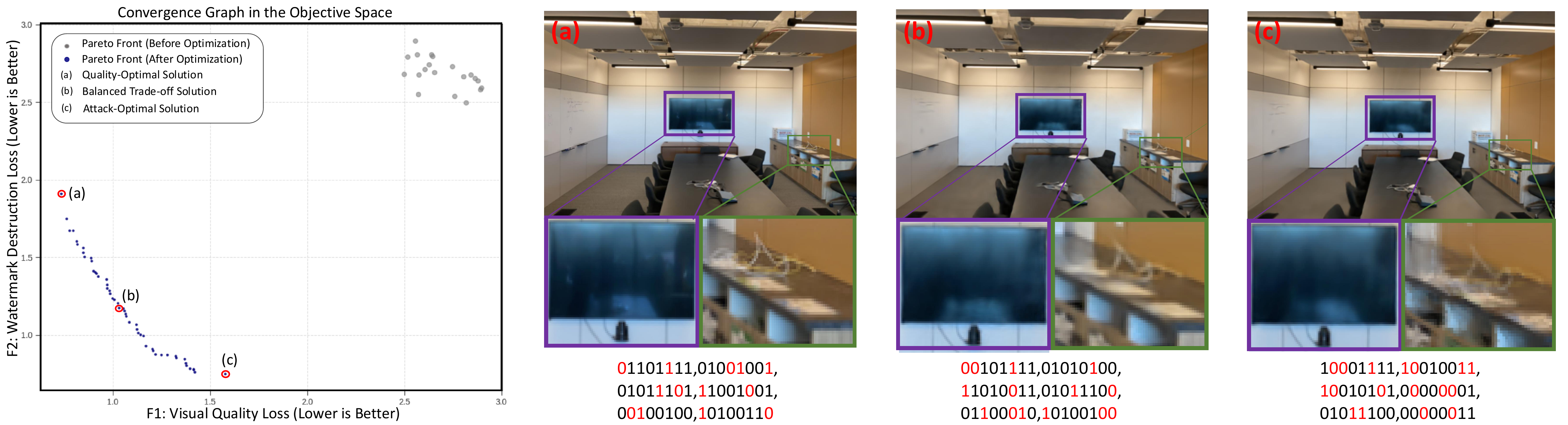}
    \caption{The GMEA attack's Pareto front, showing the trade-off between visual fidelity and attack success. The main plot highlights the optimization's improvement. Three solutions are visualized—(a) quality-optimal, (c) attack-optimal, and (b) balanced—showing their resulting render and the corrupted watermark with errors marked in red.}
\label{fig:pareto}
\end{figure*}

\begin{table*}[t!]
\centering
\scriptsize
\renewcommand{\arraystretch}{0.75}
\caption{This table quantitatively evaluates the attack's performance on 1D and 2D watermarking systems. Each row pairs a Blender scene with an LLFF scene, presenting their respective results in corresponding columns.}
\label{tab:main_results_paired}
\begin{tabular}{@{}ll ccc ccc ccc ccc@{}}
\toprule
& & \multicolumn{6}{c}{\textbf{1D Watermark (GaussianMarker)}} & \multicolumn{6}{c}{\textbf{2D Watermark (GS-Hider)}} \\
\cmidrule(lr){3-8} \cmidrule(lr){9-14}

& & \multicolumn{3}{c}{\textbf{Blender}} & \multicolumn{3}{c}{\textbf{LLFF}} & \multicolumn{3}{c}{\textbf{Blender}} & \multicolumn{3}{c}{\textbf{LLFF}} \\
\cmidrule(lr){3-5} \cmidrule(lr){6-8} \cmidrule(lr){9-11} \cmidrule(lr){12-14}

\multicolumn{1}{c}{\textbf{Scene}} & \multicolumn{1}{c}{\textbf{Method}} & BAR$\downarrow$ & WUS$\uparrow$ & IDS$\uparrow$ & BAR$\downarrow$ & WUS$\uparrow$ & IDS$\uparrow$ & SSIM$\downarrow$ & PSNR$\downarrow$ & MSE$\uparrow$ & SSIM$\downarrow$ & PSNR$\downarrow$ & MSE$\uparrow$ \\
\hline

\multirow{3}{*}{Chair / Fern} & No Attack & 99.95 & 0.1 & 0.1 & 100 & 0.0 & 0.0 & 91.18 & 15.74 & 0.026 & 97.86 & 26.43 & 0.002 \\
& GMEA (w/o $F_1$)  & 45.99 & 89.1 & 89.03 & 50.33 & 98.21 & 97.78 & 52.31 & 5.16 & 0.305 & 67.17 & 9.37 & 0.115 \\
& GMEA (ours) & 67.44 & 65.12 & 64.87 & 56.94 & 86.11 & 85.98 & 74.63 & 10.26 & 0.094 & 70.61 & 9.51 & 0.111 \\
\hline

\multirow{3}{*}{Drums / Flower} & No Attack & 99.71 & 0.58 & 0.57 & 100 & 0.0 & 0.0 & 93.12 & 18.12 & 0.015 & 96.53 & 23.67 & 0.004 \\
& GMEA (w/o $F_1$) & 45.58 & 89.96 & 90.07 & 57.92 & 84.17 & 83.9 & 52.81 & 5.03 & 0.314 & 62.48 & 8.49 & 0.141 \\
& GMEA (ours) & 65.0 & 70.0 & 69.87 & 60.83 & 78.33 & 78.25 & 75.4 & 9.92 & 0.102 & 65.42 & 11.86 & 0.065 \\
\hline

\multirow{3}{*}{Ficus / Fortress} & No Attack & 96.06 & 7.87 & 7.87 & 100 & 0.0 & 0.0 & 95.17 & 22.71 & 0.005 & 94.18 & 19.2 & 0.012 \\
& GMEA (w/o $F_1$) & 54.32 & 89.02 & 88.67 & 57.45 & 80.15 & 80.33 & 64.88 & 7.2 & 0.19 & 30.32 & 5.93 & 0.254 \\
& GMEA (ours) & 71.98 & 56.04 & 55.84 & 64.58 & 70.83 & 71.1 & 74.88 & 8.63 & 0.137 & 56.24 & 8.99 & 0.126 \\
\hline

\multirow{3}{*}{Hotdog / Horns} & No Attack & 98.61 & 2.77 & 2.73 & 100 & 0.0 & 0.0 & 95.65 & 22.71 & 0.005 & 97.86 & 26.43 & 0.002 \\
& GMEA (w/o $F_1$) & 58.16 & 82.77 & 82.85 & 60.36 & 77.81 & 77.63 & 60.36 & 6.63 & 0.217 & 60.09 & 8.7 & 0.135 \\
& GMEA (ours) & 58.21 & 83.17 & 83.11 & 61.98 & 76.04 & 75.9 & 73.7 & 9.66 & 0.108 & 75.75 & 12.68 & 0.053 \\
\hline

\multirow{3}{*}{Lego / Leaves} & No Attack & 99.99 & 0.02 & 0.02 & 100 & 0.0 & 0.0 & 94.33 & 21.19 & 0.007 & 98.54 & 29.93 & 0.001 \\
& GMEA (w/o $F_1$) & 45.19 & 88.21 & 88.03 & 60.13 & 78.21 & 78.15 & 68.98 & 6.76 & 0.211 & 71.73 & 11.39 & 0.072 \\
& GMEA (ours) & 68.53 & 67.35 & 65.88 & 68.75 & 62.5 & 62.47 & 74.45 & 10.24 & 0.094 & 80.2 & 13.26 & 0.047 \\
\hline

\multirow{3}{*}{Materials / Trex} & No Attack & 99.43 & 1.15 & 1.13 & 100 & 0.0 & 0.0 & 91.67 & 17.75 & 0.016 & 97.44 & 28.03 & 0.001 \\
& GMEA (w/o $F_1$) & 55.05 & 83.69 & 83.75 & 59.45 & 77.25 & 77.41 & 60.22 & 6.25 & 0.237 & 64.96 & 9.15 & 0.121 \\
& GMEA (ours) & 63.98 & 71.96 & 71.93 & 65.63 & 68.75 & 68.4 & 79.07 & 10.33 & 0.092 & 69.88 & 10.95 & 0.081 \\
\hline

\multirow{3}{*}{Mic / Room} & No Attack & 99.21 & 1.58 & 1.57 & 100 & 0.0 & 0.0 & 92.7 & 18.64 & 0.013 & 98.28 & 29.74 & 0.001 \\
& GMEA (w/o $F_1$) & 56.22 & 86.44 & 86.46 & 66.14 & 65.43 & 65.58 & 32.52 & 2.19 & 0.603 & 70.4 & 9.59 & 0.109 \\
& GMEA (ours) & 67.19 & 65.58 & 65.51 & 71.53 & 56.94 & 56.62 & 77.79 & 8.95 & 0.127 & 75.07 & 11.26 & 0.074 \\
\hline

\multirow{3}{*}{Ship / Orchids} & No Attack & 99.4 & 1.21 & 1.19 & 100 & 0.0 & 0.0 & 95.54 & 21.41 & 0.007 & 98.68 & 31.89 & 0.001 \\
& GMEA (w/o $F_1$) & 56.65 & 84.25 & 84.32 & 53.18 & 93.84 & 94.01 & 63.47 & 6.56 & 0.22 & 61.18 & 8.06 & 0.156 \\
& GMEA (ours) & 64.9 & 70.12 & 70.04 & 55.95 & 88.1 & 88.19 & 72.56 & 8.67 & 0.135 & 67.72 & 9.15 & 0.121 \\

\bottomrule
\end{tabular}
\end{table*}

\subsubsection{Reconstructing the Adversarial Model.}
The final step is to reconstruct the complete adversarial model, $\mathbf{G}_{adv}$. Since our group-based strategy operates on disjoint sets of Gaussian kernels, this reconstruction is a straightforward union of the $k$ optimized sub-models:
\begin{equation}
\mathbf{G}_{adv} = \bigcup_{i=1}^{k} \mathbf{G}^{(i)}_{adv}.
\end{equation}

The resulting model $\mathbf{G}_{adv}$ aggregates all modifications from the independent optimization runs and represents the final output of our attack framework.

\section{Evaluation and Results}

\subsection{Experimental Settings}

\subsubsection{Model and dataset.}
To demonstrate GMEA's versatility, we target representative systems from two distinct categories of 3DGS watermarking: GaussianMarker \cite{huang2024gaussianmarker} for 1D bitstream watermarks and GS-Hider \cite{NEURIPS2024_59091e82} for 2D image watermarks. The evaluation was performed on two datasets: Blender dataset \cite{mildenhall2021nerf}, comprising objects without backgrounds, and the more challenging LLFF dataset \cite{mildenhall2019local}, which features complex real-world scenes.



\subsubsection{Evaluation metrics.}

We assess our framework's performance based on two criteria: visual fidelity and watermark removal efficacy. Visual quality is quantified using standard image metrics: Peak Signal-to-Noise Ratio (PSNR), the Structural Similarity Index Measure (SSIM), and Mean Squared Error (MSE) \cite{RN912}. For evaluating 1D watermark removal, we use the standard Bit Accuracy Rate (BAR) and our proposed Watermark Uncertainty Score (WUS) and Information Destruction Score (IDS). Detailed definitions for these metrics, along with further experimental settings and results, \underline{are deferred to the Appendix.}


\subsection{Experiment Results}

\subsubsection{Attack Performance Evaluation.}
We assessed our GMEA framework's effectiveness through extensive experiments on 1D and 2D watermarking systems, as shown in \Cref{tab:main_results_paired}. Our full GMEA attack significantly disrupts the near-perfect watermark extraction of the \textit{No Attack} baseline in the 1D watermarking system, reducing the Bit Accuracy Rate (BAR) to an average of approximately 65\% and substantially increasing the Watermark Uncertainty Score (WUS) and Information Destruction Score (IDS). This renders the extracted bitstream highly unreliable.

In the 2D watermarking system, our full GMEA method's success is evident in the degradation of the extracted watermark image's quality, with a drastic drop in both SSIM and PSNR values compared to the $\textit{No Attack}$ baseline. This severe degradation, detailed in \Cref{tab:main_results_paired}, confirms our attack effectively renders the 2D visual watermark unrecognizable. These findings validate our GMEA's capability to successfully compromise the detectability of different mainstream watermarking schemes.



\begin{figure}
    \centering
    \includegraphics[width=1\linewidth]{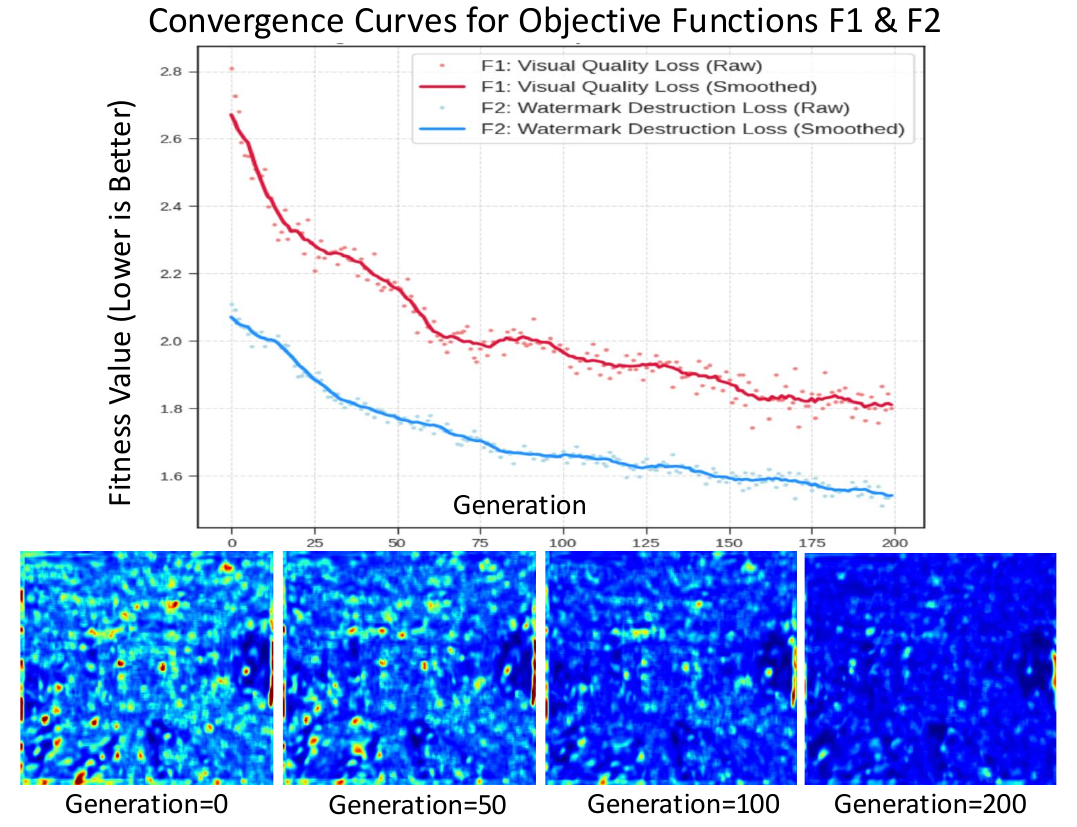}
    \caption{Visualization of GMEA's optimization. The top plot shows the convergence of objectives $F_1$ and $F_2$. The bottom panels show the feature map becoming uniform, which visually confirms the destruction of the watermark.}
\label{fig:convergence_and_features}
\end{figure}

\subsubsection{Ablation Study.}

To further analyze the components of our framework, we conducted an ablation study by removing the visual quality objective ($F_1$) and creating a single-objective variant, GMEA (w/o $F_1$), which solely optimizes for watermark destruction ($F_2$). As shown in \Cref{tab:main_results_paired}, this ablation variant exhibits even more potent attack capabilities.

For the 1D watermark, GMEA (w/o $F_1$) achieves a BAR closer to 50\% (the theoretical value for random guessing) and higher WUS/IDS scores than our full two-objective method GMEA. For instance, in the Materials / Trex scene, the WUS score for GMEA (w/o $F_1$) attack reaches 83.69, compared to 71.96 for the full GMEA. For the 2D watermark, the GMEA (w/o $F_1$) attack results in a significantly lower SSIM/PSNR  for the extracted watermark image, indicating more severe corruption. This demonstrates that by focusing exclusively on maximizing watermark destruction without the constraint of preserving visual fidelity, the attack can achieve a higher degree of watermark removal. However, it completely disregards maintaining the visual quality of the 3DGS model, resulting in the 3DGS being unusable after watermark removal. A detailed analysis of the resulting visual degradation is presented in the next section.

\subsubsection{Justification for the Multi-objective Approach.}
While the single-objective attack GMEA (w/o $F_1$) offers superior watermark destruction, it comes at the unacceptable cost of visual degradation to the 3DGS model. This trade-off is qualitatively illustrated in \Cref{fig:qualitative_comparison}. As the figure shows, although the watermark extracted by GMEA (w/o $F_1$) is more severely distorted, the rendered image is also riddled with visual artifacts, unlike the pristine render from our full GMEA.

The quantitative data presented in \Cref{tab:ablation_fidelity_restructured} corroborates this visual evidence. Our full GMEA method maintains high SSIM (above 95\%) and PSNR values, indicating minimal distortion. In contrast, the GMEA (w/o $F_1$) variant causes a drastic drop in these metrics. Since an attack that destroys an asset's visual value defeats the purpose of the theft, this ablation study validates the necessity of our multi-objective formulation, which effectively balances potent watermark removal with the preservation of high visual fidelity.




\begin{table}[t!]
\centering
\scriptsize
\renewcommand{\arraystretch}{0.7}
\caption{Assessing the visual distortion of 3DGS introduced by our attack via an ablation study. }
\label{tab:ablation_fidelity_restructured}
\begin{tabular}{@{}l l ccc@{}}
\toprule
\textbf{Watermarking Target} & \textbf{Attack Method} & \textbf{SSIM}$\uparrow$ & \textbf{PSNR}$\uparrow$ & \textbf{MSE}$\downarrow$ \\
\midrule
\multirow{2}{*}{1D (GaussianMarker)} 
& GMEA (w/o $F_1$) & 67.19 & 20.31 & 0.0112 \\
& GMEA (ours) & \textbf{95.13} & \textbf{30.66} & \textbf{0.0011} \\
\midrule
\multirow{2}{*}{2D (GS-Hider)} 
& GMEA (w/o $F_1$) & 60.51 & 17.19 & 0.0263 \\
& GMEA (ours) & \textbf{98.22} & \textbf{37.76} & \textbf{0.0002} \\
\bottomrule
\end{tabular}
\end{table}

\subsubsection{Optimization Dynamics and Convergence.}

\Cref{fig:convergence_and_features} visualizes the operational dynamics of our GMEA. The top graph shows the steady convergence of both visual quality loss ($F_1$) and watermark destruction loss ($F_2$), demonstrating the effectiveness of our multi-objective evolutionary algorithm. The bottom panels qualitatively validate our watermark destruction objective ($F_2$) by displaying visualizations of the intermediate convolutional feature map. At Generation 0, the feature map exhibits distinct spatial patterns that are essential for watermark decoders. As the optimization progresses, these patterns are suppressed, and by Generation 200, the feature map is largely homogeneous, erasing its discriminative features. Since any potential downstream decoder must  rely on these upstream convolutional features, this flattening effectively blinds the entire watermark extraction pipeline. This visual collapse of feature information directly corresponds to the convergence of  the $F_2$ curve, proving that minimizing feature variance is a successful and universal indirect attack strategy in a black-box setting.

\subsubsection{Pareto Front Analysis and Solution Trade-offs.}

A key strength of our multi-objective approach is its ability to find a set of optimal solutions representing the trade-offs between conflicting objectives. \Cref{fig:pareto} analyzes the trade-off between attack effectiveness (low $F_2$) and visual fidelity (low $F_1$). The main plot shows a significant improvement, with solutions evolving from a suboptimal initial state (top-right, gray) to a superior final Pareto front (bottom-left, blue).

For a more granular analysis of the solution space, we visualize three representative solutions from the final front. The Attack-Optimal Solution (c) achieves the most effective watermark corruption at the cost of slight visual artifacts. Conversely, the Quality-Optimal Solution (a) yields the highest visual fidelity but with a less damaged watermark. The Balanced Trade-off Solution (b) presents an ideal compromise, achieving significant watermark disruption with negligible visual distortion. This analysis highlights GMEA's superiority: it discovers a range of effective attack solutions and offers the flexibility to select one based on the desired balance between efficacy and fidelity.

\begin{table}[t!]
\centering
\small
\renewcommand{\arraystretch}{0.6} 
\caption{Time-memory trade-off analysis for our group-based strategy. The table shows the computational time and peak GPU memory required to reach a fixed target fitness ($F_1 \approx 1.9, F_2 \approx 1.7$) as the number of groups ($k$) varies.}
\label{tab:group_cost_analysis}
\begin{tabular}{@{}ccc@{}}
\toprule
\textbf{Groups ($k$)} & \textbf{Time (min)} & \textbf{GPU Memory (GB)} \\
\midrule
1  & 74.2 & 24.0 \\
5  & 32.5 & 33.6 \\
10 & 23.7 & 43.2 \\
20 & 20.1 & 55.4 \\
50 & 18.5 & 84.9 \\
\bottomrule
\end{tabular}
\end{table}

\subsubsection{Analysis of the Group-Based Strategy.}
We analyzed the impact of the number of groups ($k$) on optimization efficiency, with results shown in \Cref{tab:group_cost_analysis}. The data reveals a clear time-memory trade-off: increasing $k$ accelerates convergence by parallelizing the search, but at the cost of higher peak GPU memory.  Considering this trade-off, we identify $k=10$ as a balanced setting for our main experiments, as it provides a significant speedup while maintaining a manageable memory footprint.  More analysis is in the Appendix.




\section{Conclusion}

This paper introduced GMEA, the first universal black-box framework for assessing the robustness of 3DGS watermarking systems. By formulating the attack as a group-based multi-objective optimization problem, GMEA effectively balances watermark removal with visual fidelity preservation, using a novel feature-variance objective to operate without detector knowledge. Our experiments show that GMEA effectively compromises both 1D and 2D watermarking schemes while maintaining high visual fidelity.


\bibliography{aaai2026}

\begin{thebibliography}{33}
\providecommand{\natexlab}[1]{#1}

\bibitem[{Bao et~al.(2025)Bao, Ding, Huo, Liu, Li, Li, Gao, and Luo}]{10870258}
Bao, Y.; Ding, T.; Huo, J.; Liu, Y.; Li, Y.; Li, W.; Gao, Y.; and Luo, J. 2025.
\newblock 3D Gaussian Splatting: Survey, Technologies, Challenges, and Opportunities.
\newblock \emph{IEEE Transactions on Circuits and Systems for Video Technology}, 35(7): 6832--6852.

\bibitem[{Chen et~al.(2025{\natexlab{a}})Chen, Wang, Zhu, Lai, and Xie}]{chen2025guardsplat}
Chen, Z.; Wang, G.; Zhu, J.; Lai, J.; and Xie, X. 2025{\natexlab{a}}.
\newblock GuardSplat: Efficient and Robust Watermarking for 3D Gaussian Splatting.
\newblock In \emph{Processings of the IEEE/CVF Conference on Computer Vision and Pattern Recognition}, 16325--16335.

\bibitem[{Chen et~al.(2025{\natexlab{b}})Chen, Yang, Huang, de~Lutio, Esturo, Ivanovic, Litany, Gojcic, Fidler, Pavone, Song, and Wang}]{chen2025omnire}
Chen, Z.; Yang, J.; Huang, J.; de~Lutio, R.; Esturo, J.~M.; Ivanovic, B.; Litany, O.; Gojcic, Z.; Fidler, S.; Pavone, M.; Song, L.; and Wang, Y. 2025{\natexlab{b}}.
\newblock OmniRe: Omni Urban Scene Reconstruction.
\newblock In \emph{International Conference on Learning Representations}.

\bibitem[{Cox et~al.(1997)Cox, Kilian, Leighton, and Shamoon}]{cox1997secure}
Cox, I.~J.; Kilian, J.; Leighton, F.~T.; and Shamoon, T. 1997.
\newblock Secure spread spectrum watermarking for multimedia.
\newblock \emph{IEEE Transactions on Image Processing}, 6(12): 1673--1687.

\bibitem[{Deb et~al.(2002)Deb, Pratap, Agarwal, and Meyarivan}]{deb2002fast}
Deb, K.; Pratap, A.; Agarwal, S.; and Meyarivan, T. 2002.
\newblock A fast and elitist multiobjective genetic algorithm: NSGA-II.
\newblock \emph{IEEE Transactions on Evolutionary Computation}, 6(2): 182--197.

\bibitem[{Gu\'edon and Lepetit(2024)}]{guedon2023sugar}
Gu\'edon, A.; and Lepetit, V. 2024.
\newblock SuGaR: Surface-Aligned Gaussian Splatting for Efficient 3D Mesh Reconstruction and High-Quality Mesh Rendering.
\newblock In \emph{Processings of the IEEE/CVF Conference on Computer Vision and Pattern Recognition}, 5354--5363.

\bibitem[{Hong, Jiang, and Yen(2023)}]{hong2023improving}
Hong, H.; Jiang, M.; and Yen, G.~G. 2023.
\newblock Improving performance insensitivity of large-scale multiobjective optimization via Monte Carlo tree search.
\newblock \emph{IEEE Transactions on Cybernetics}, 54(3): 1816--1827.

\bibitem[{Huang et~al.(2025)Huang, Li, Cheung, Cheung, See, and Wan}]{huang2024gaussianmarker}
Huang, X.; Li, R.; Cheung, Y.-m.; Cheung, K.~C.; See, S.; and Wan, R. 2025.
\newblock Gaussianmarker: Uncertainty-aware copyright protection of 3d gaussian splatting.
\newblock In \emph{Processings of Neural Information Processing Systems}.

\bibitem[{Jang et~al.(2025)Jang, Park, Yang, Ko, Choo, and Kim}]{jang20253dgsw3dgaussiansplatting}
Jang, Y.; Park, H.; Yang, F.; Ko, H.; Choo, E.; and Kim, S. 2025.
\newblock 3D-GSW: 3D Gaussian Splatting for Robust Watermarking.
\newblock arXiv:2409.13222.

\bibitem[{Kerbl et~al.(2023)Kerbl, Kopanas, Leimkhler, and Drettakis}]{kerbl3Dgaussians}
Kerbl, B.; Kopanas, G.; Leimkhler, T.; and Drettakis, G. 2023.
\newblock 3D Gaussian Splatting for Real-Time Radiance Field Rendering.
\newblock \emph{ACM Transactions on Graphics}, 42(4).

\bibitem[{Liang et~al.(2023)Liang, Ban, Yu, Qu, Qiao, Yue, Chen, and Tan}]{9723472}
Liang, J.; Ban, X.; Yu, K.; Qu, B.; Qiao, K.; Yue, C.; Chen, K.; and Tan, K.~C. 2023.
\newblock A Survey on Evolutionary Constrained Multiobjective Optimization.
\newblock \emph{IEEE Transactions on Evolutionary Computation}, 27(2): 201--221.

\bibitem[{Lu et~al.(2020)Lu, Jia, Wang, Li, Chai, Carin, and Velipasalar}]{lu2020enhancing}
Lu, Y.; Jia, Y.; Wang, J.; Li, B.; Chai, W.; Carin, L.; and Velipasalar, S. 2020.
\newblock Enhancing cross-task black-box transferability of adversarial examples with dispersion reduction.
\newblock In \emph{Proceedings of the IEEE/CVF conference on Computer Vision and Pattern Recognition}, 940--949.

\bibitem[{MacQueen(1967)}]{MacQueen1967SomeMF}
MacQueen, J. 1967.
\newblock Some methods for classification and analysis of multivariate observations.

\bibitem[{Mildenhall et~al.(2019)Mildenhall, Srinivasan, Ortiz-Cayon, Kalantari, Ramamoorthi, Ng, and Kar}]{mildenhall2019local}
Mildenhall, B.; Srinivasan, P.~P.; Ortiz-Cayon, R.; Kalantari, N.~K.; Ramamoorthi, R.; Ng, R.; and Kar, A. 2019.
\newblock Local light field fusion: Practical view synthesis with prescriptive sampling guidelines.
\newblock \emph{ACM Transactions on Graphics}, 38(4): 1--14.

\bibitem[{Mildenhall et~al.(2021)Mildenhall, Srinivasan, Tancik, Barron, Ramamoorthi, and Ng}]{mildenhall2021nerf}
Mildenhall, B.; Srinivasan, P.~P.; Tancik, M.; Barron, J.~T.; Ramamoorthi, R.; and Ng, R. 2021.
\newblock Nerf: Representing scenes as neural radiance fields for view synthesis.
\newblock \emph{Communications of the ACM}, 65(1): 99--106.

\bibitem[{Papernot et~al.(2017)Papernot, McDaniel, Goodfellow, Jha, Celik, and Swami}]{papernot2017practical}
Papernot, N.; McDaniel, P.; Goodfellow, I.; Jha, S.; Celik, Z.~B.; and Swami, A. 2017.
\newblock Practical black-box attacks against machine learning.
\newblock In \emph{ACM on Asia Conference on Computer and Communications Security}, 506--519.

\bibitem[{Setiadi(2021)}]{RN912}
Setiadi, D. R. I.~M. 2021.
\newblock PSNR vs SSIM: imperceptibility quality assessment for image steganography.
\newblock \emph{Multimedia Tools and Applications}, 80(6): 8423--8444.

\bibitem[{Tu et~al.(2025)Tu, Radl, Steiner, Steinberger, Kerbl, and de~la Torre}]{Tu2025VRSplat}
Tu, X.; Radl, L.; Steiner, M.; Steinberger, M.; Kerbl, B.; and de~la Torre, F. 2025.
\newblock VRSplat: Fast and Robust Gaussian Splatting for Virtual Reality.
\newblock \emph{Processings of the ACM on Computer Graphics and Interactive Techniques}, 8(1).

\bibitem[{Van~Gansbeke et~al.(2020)Van~Gansbeke, Vandenhende, Georgoulis, Proesmans, and Van~Gool}]{10.1007/978-3-030-58607-2_16}
Van~Gansbeke, W.; Vandenhende, S.; Georgoulis, S.; Proesmans, M.; and Van~Gool, L. 2020.
\newblock Scan: Learning to classify images without labels.
\newblock In \emph{Processings of European Conference on Computer Vision}, 268--285. Springer.

\bibitem[{Wang et~al.(2021)Wang, Wang, Wang, and Zhang}]{wang2021survey}
Wang, J.; Wang, Y.; Wang, H.; and Zhang, J. 2021.
\newblock A survey on evolutionary computation for adversarial machine learning.
\newblock \emph{IEEE Transactions on Evolutionary Computation}, 26(5): 994--1009.

\bibitem[{Wang et~al.(2024)Wang, Zeng, Lin, Jiang, and Tan}]{wang2024generating}
Wang, Z.; Zeng, Q.; Lin, W.; Jiang, M.; and Tan, K.~C. 2024.
\newblock Generating Diagnostic and Actionable Explanations for Fair Graph Neural Networks.
\newblock In \emph{Processings of the AAAI Conference on Artificial Intelligence}, volume~38, 21690--21698.

\bibitem[{Wu et~al.(2024{\natexlab{a}})Wu, Yi, Fang, Xie, Zhang, Wei, Liu, Tian, and Wang}]{Wu_2024_CVPR}
Wu, G.; Yi, T.; Fang, J.; Xie, L.; Zhang, X.; Wei, W.; Liu, W.; Tian, Q.; and Wang, X. 2024{\natexlab{a}}.
\newblock 4D Gaussian Splatting for Real-Time Dynamic Scene Rendering.
\newblock In \emph{Processings of the IEEE/CVF Conference on Computer Vision and Pattern Recognition}, 20310--20320.

\bibitem[{Wu et~al.(2024{\natexlab{b}})Wu, Yuan, Zhang, Yang, Cao, Yan, and Gao}]{10897713}
Wu, T.; Yuan, Y.-J.; Zhang, L.-X.; Yang, J.; Cao, Y.-P.; Yan, L.-Q.; and Gao, L. 2024{\natexlab{b}}.
\newblock Recent advances in 3D Gaussian splatting.
\newblock \emph{Computational Visual Media}, 10(4): 613--642.

\bibitem[{Xie et~al.(2024)Xie, Zong, Qiu, Li, Feng, Yang, and Jiang}]{xie2023physgaussian}
Xie, T.; Zong, Z.; Qiu, Y.; Li, X.; Feng, Y.; Yang, Y.; and Jiang, C. 2024.
\newblock PhysGaussian: Physics-Integrated 3D Gaussians for Generative Dynamics.
\newblock In \emph{Proceedings of the IEEE/CVF Conference on Computer Vision and Pattern Recognition}, 4389--4398.

\bibitem[{Yi et~al.(2024)Yi, Fang, Wang, Wu, Xie, Zhang, Liu, Tian, and Wang}]{10655860}
Yi, T.; Fang, J.; Wang, J.; Wu, G.; Xie, L.; Zhang, X.; Liu, W.; Tian, Q.; and Wang, X. 2024.
\newblock GaussianDreamer: Fast Generation from Text to 3D Gaussians by Bridging 2D and 3D Diffusion Models.
\newblock In \emph{IEEE/CVF Conference on Computer Vision and Pattern Recognition}, 6796--6807.

\bibitem[{Zeng et~al.(2024)Zeng, Wang, Cheung, and Jiang}]{zeng2024ask}
Zeng, Q.; Wang, Z.; Cheung, Y.-m.; and Jiang, M. 2024.
\newblock Ask, attend, attack: An effective decision-based black-box targeted attack for image-to-text models.
\newblock In \emph{Processings of Neural Information Processing Systems}, volume~37, 105819--105847.

\bibitem[{Zhang et~al.(2020)Zhang, Wang, Liu, and Zhong}]{zhang2020survey}
Zhang, R.-Z.; Wang, S.-Z.; Liu, T.-C.; and Zhong, S.-M. 2020.
\newblock A survey on adversarial attacks on deep-learning based steganography.
\newblock \emph{IEEE Access}, 8: 189518--189537.

\bibitem[{Zhang et~al.(2024{\natexlab{a}})Zhang, Yu, Wu, Feng, Zheng, Snavely, Wu, and Freeman}]{zhang2024physdreamer}
Zhang, T.; Yu, H.-X.; Wu, R.; Feng, B.~Y.; Zheng, C.; Snavely, N.; Wu, J.; and Freeman, W.~T. 2024{\natexlab{a}}.
\newblock Physdreamer: Physics-based interaction with 3d objects via video generation.
\newblock In \emph{Processings of European Conference on Computer Vision}, 388--406. Springer.

\bibitem[{Zhang et~al.(2024{\natexlab{b}})Zhang, Meng, Li, Xu, Zhang, and Zhang}]{NEURIPS2024_59091e82}
Zhang, X.; Meng, J.; Li, R.; Xu, Z.; Zhang, Y.; and Zhang, J. 2024{\natexlab{b}}.
\newblock GS-Hider: Hiding Messages into 3D Gaussian Splatting.
\newblock In Globerson, A.; Mackey, L.; Belgrave, D.; Fan, A.; Paquet, U.; Tomczak, J.; and Zhang, C., eds., \emph{Processings of Neural Information Processing Systems}, volume~37, 49780--49805.

\bibitem[{Zhang et~al.(2018)Zhang, Tian, Cheng, and Jin}]{7544478}
Zhang, X.; Tian, Y.; Cheng, R.; and Jin, Y. 2018.
\newblock A Decision Variable Clustering-Based Evolutionary Algorithm for Large-Scale Many-Objective Optimization.
\newblock \emph{IEEE Transactions on Evolutionary Computation}, 22(1): 97--112.

\bibitem[{Zhao et~al.(2024)Zhao, Zhang, Su, Vasan, Grishchenko, Kruegel, Vigna, Wang, and Li}]{zhao2024invisible}
Zhao, X.; Zhang, K.; Su, Z.; Vasan, S.; Grishchenko, I.; Kruegel, C.; Vigna, G.; Wang, Y.-X.; and Li, L. 2024.
\newblock Invisible image watermarks are provably removable using generative ai.
\newblock In \emph{Processings of Neural Information Processing Systems}, volume~37, 8643--8672.

\bibitem[{Zhu et~al.(2025)Zhu, Liang, Chang, Deng, Lu, Yang, Zhang, and Zhang}]{10.5555/3737916.3741145}
Zhu, R.; Liang, Y.; Chang, H.; Deng, J.; Lu, J.; Yang, W.; Zhang, T.; and Zhang, Y. 2025.
\newblock MotionGS: exploring explicit motion guidance for deformable 3D Gaussian splatting.
\newblock In \emph{Processings of Neural Information Processing Systems}. Red Hook, NY, USA.
\newblock ISBN 9798331314385.

\bibitem[{Zhu et~al.(2024)Zhu, Wang, Kong, Kong, and Wang}]{zhu20243dgaussiansplattingrobotics}
Zhu, S.; Wang, G.; Kong, X.; Kong, D.; and Wang, H. 2024.
\newblock 3D Gaussian Splatting in Robotics: A Survey.
\newblock arXiv:2410.12262.

\end{thebibliography}

\clearpage
\appendix
\section*{Appendix}

\setcounter{secnumdepth}{1}

\begin{figure*}
    \centering
    \includegraphics[width=1\linewidth]{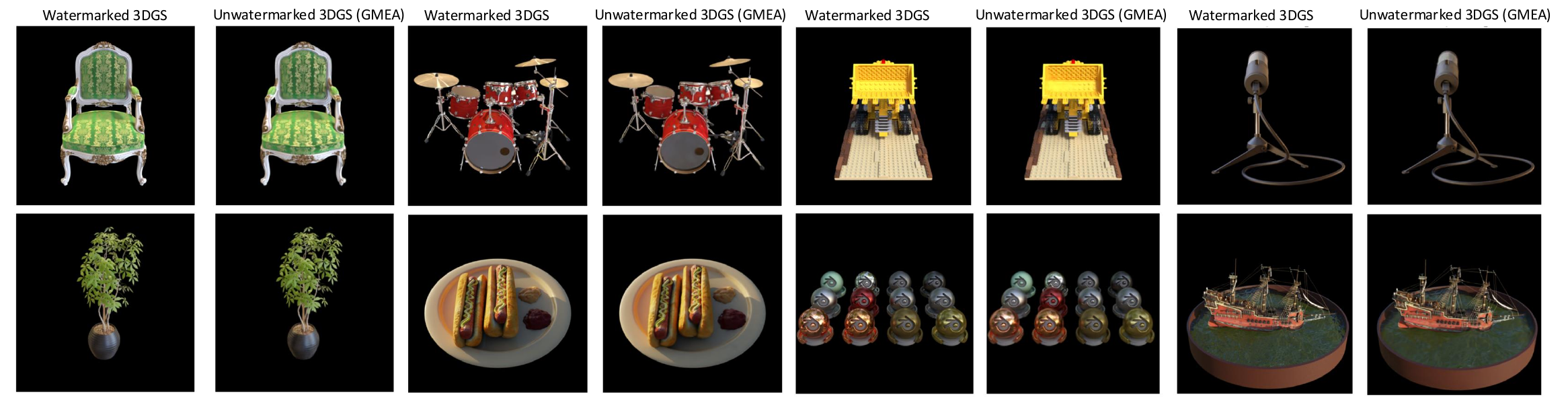}
    \caption{Qualitative comparison on the \textbf{Blender dataset}. Each pair displays the rendered image from the original watermarked 3DGS model (left) alongside the render from the model after being processed by our GMEA framework (right). The high degree of visual similarity demonstrates that our attack successfully removes the watermark while preserving excellent visual fidelity, making the modifications virtually imperceptible.}
    \label{fig:qualitative_blender}
\end{figure*}

\section{Overview}
This appendix provides supplementary material to support and elaborate on the findings presented in our main paper, "Fading the Digital Ink: A Universal Black-Box Attack Framework for 3DGS Watermarking Systems." The goal is to offer greater detail and transparency regarding our methodology, implementation, and results. The structure of this appendix is as follows:

\begin{itemize}
    \item \textbf{Reproducibility Details:} A comprehensive guide to reproducing our work. This includes the detailed experimental settings (hyperparameters, hardware, software, watermark specifications, and evaluation metrics) and the formal pseudocode for our GMEA framework and its sub-components.

    \item \textbf{Theoretical Justification:} The complete mathematical proof for our watermark destruction objective ($F_2$), which formally establishes the relationship between minimizing feature variance and erasing information content.

    \item \textbf{Supplementary Experiments:} A collection of additional results that were omitted from the main paper due to space constraints. This includes extensive qualitative comparisons on multiple datasets, a visual analysis of 2D watermark corruption, a full quantitative breakdown of visual fidelity across all attack variations, and an in-depth ablation study on the impact of population size.

    \item \textbf{Discussion:} A broader discussion of our work, including an acknowledgment of the current limitations (such as computational cost) and an outline of promising directions for future research.
\end{itemize}

We believe these materials offer a comprehensive resource for readers interested in the technical details and reproducibility of our work.

\begin{figure}[h!]
    \centering
    \includegraphics[width=1\linewidth]{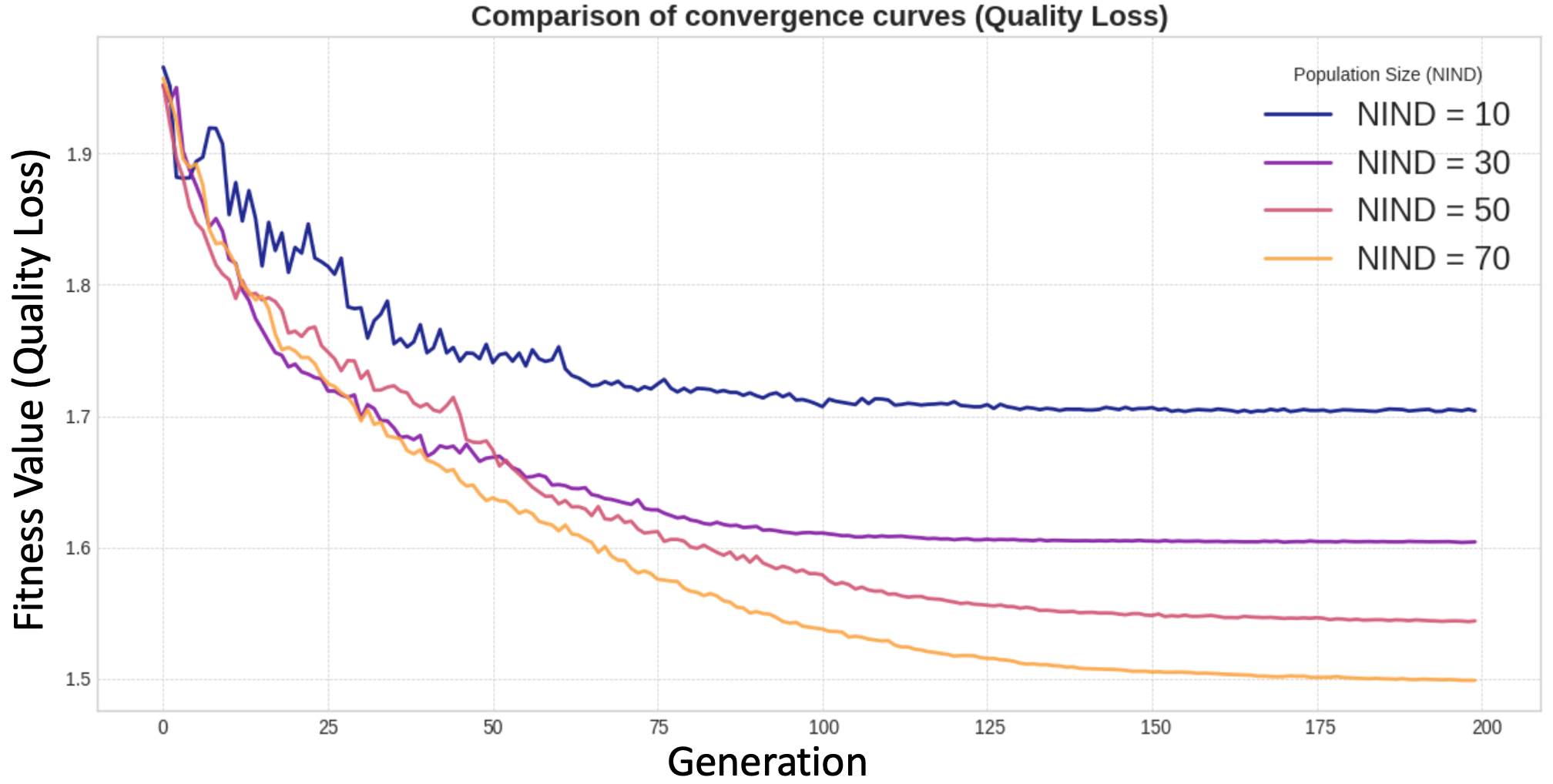}
    \caption{Convergence of Objective 1 (Quality Loss) vs. \textbf{Generation}. This plot shows that larger population sizes (e.g., NIND=70) achieve a lower (better) fitness value over 200 generations. However, this view does not account for the differing time costs.}
    \label{fig:convergence_f1_vs_gen}
\end{figure}

\begin{figure}[h!]
    \centering
    \includegraphics[width=1\linewidth]{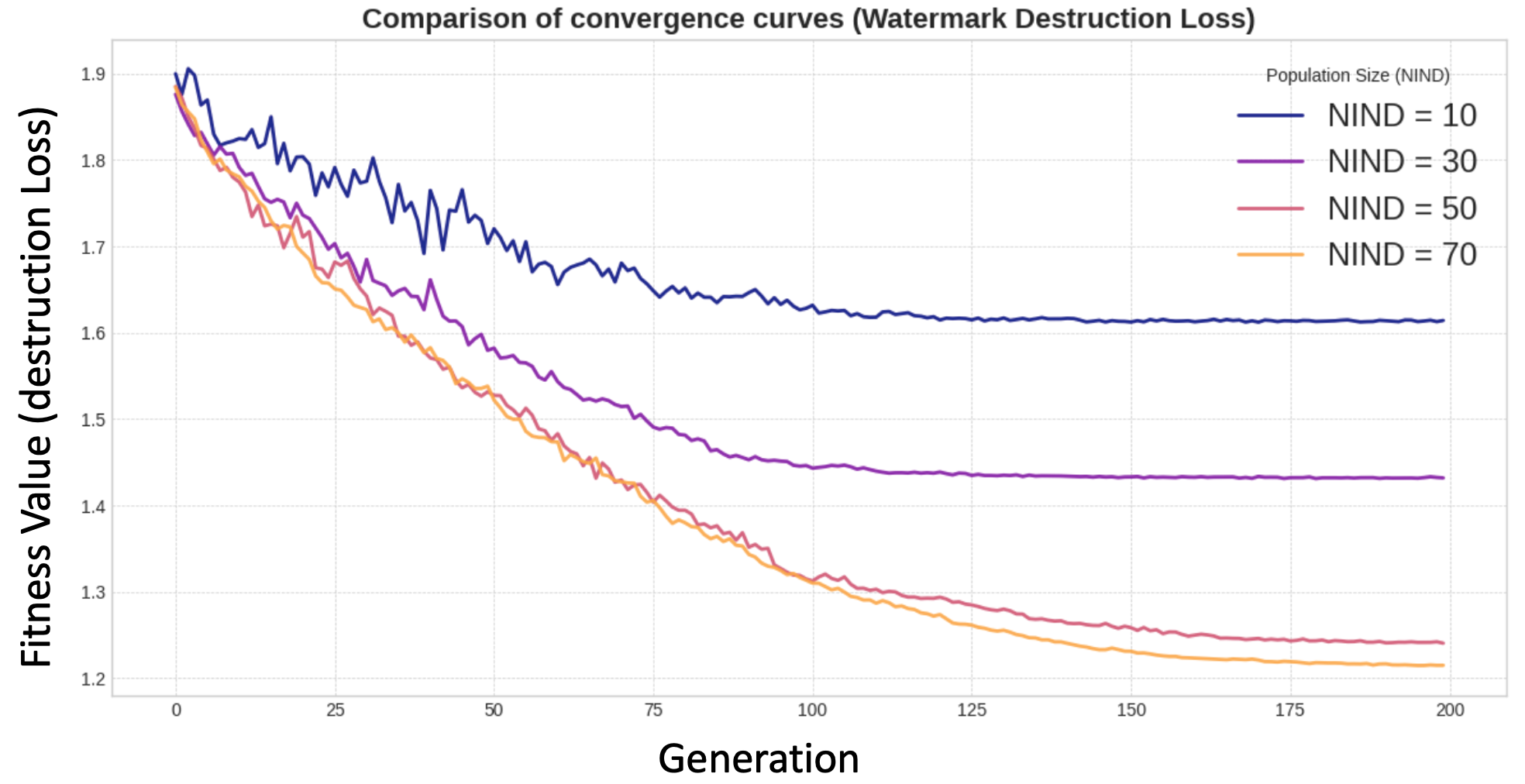}
    \caption{Convergence of Objective 2 (Watermark Destruction) vs. \textbf{Generation}. Similar to F1, larger populations show better convergence per generation, finding solutions that more effectively destroy the watermark by the end of the run.}
    \label{fig:convergence_f2_vs_gen}
\end{figure}





\begin{table}[h!]
\centering
\caption{Visual fidelity of the attacked 3DGS models (1D watermark, Blender dataset) compared to the original watermarked models. Our full GMEA method maintains high visual quality, as shown by the high SSIM and PSNR, and low RMSE. The arrows indicate the desired direction for each metric.}
\label{tab:visual_fidelity_1d_blender}
\begin{tabular}{@{}lccc@{}}
\toprule
\textbf{Scene} & \textbf{SSIM} $\uparrow$ & \textbf{PSNR} $\uparrow$ & \textbf{RMSE} $\downarrow$ \\
\midrule
Chair          & 95.48 & 29.05 & 0.0363 \\
Drums          & 94.20 & 26.77 & 0.0489 \\
Ficus          & 97.25 & 32.73 & 0.0258 \\
Hotdog         & 96.82 & 33.27 & 0.0242 \\
Lego           & 94.74 & 30.15 & 0.0347 \\
Materials      & 95.57 & 30.92 & 0.0317 \\
Mic            & 96.87 & 32.39 & 0.0263 \\
Ship           & 90.13 & 30.00 & 0.0342 \\
\midrule
\textbf{Average} & \textbf{95.13} & \textbf{30.66} & \textbf{0.0328} \\
\bottomrule
\end{tabular}
\end{table}

\section{Reproducibility}

Our \textbf{source code} and \textbf{data} are included in the supplemental material, and we will publish the code on GitHub after the paper is accepted to ensure full reproducibility.

\subsection{Experimental Settings}
To ensure the transparency and reproducibility of our work, this section provides a comprehensive overview of the experimental environment and the hyperparameter configurations used for our GMEA framework. To ensure statistical reliability and account for the stochastic nature of the evolutionary algorithm, all reported quantitative results represent \underline{the average of 5 independent runs}, each with a different random seed.

\subsubsection{Watermark Details}
\begin{figure}[h!]
\centering
\includegraphics[width=0.6\linewidth]{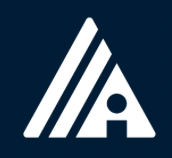}
\caption{The 2D image watermark (AAAI logo) used for all experiments targeting the GS-Hider watermarking system.}
\label{fig:2d_watermark_logo}
\end{figure}

To ensure our experiments are fully reproducible, this section details the specific 1D and 2D watermarks used. For all experiments involving 1D watermarking (targeting the GaussianMarker method), we used a fixed 48-bit binary string as the copyright message to ensure consistency across different scenes and attack evaluations. The specific bitstream embedded was:
\begin{center}
\texttt{11101011, 01010000,01010111, \\ 01001101, 01000100, 00100111}
\end{center}
For the 2D watermarking experiments (targeting the GS-Hider method), the AAAI logo served as the embedded image watermark, as visualized in \Cref{fig:2d_watermark_logo}.

\subsubsection{Evaluation Metrics}
We assess our framework's performance based on two primary criteria: the visual fidelity of the attacked 3DGS model and the efficacy of watermark removal. Visual quality is quantified using standard image metrics: PSNR, SSIM(\%), and MSE, where higher PSNR and SSIM values, alongside a lower MSE, indicate better preservation of visual fidelity. For 1D bitstream watermarks, we measure removal efficacy with three metrics. The first is the standard \textbf{Bit Accuracy Rate (BAR)}, or the proportion of correctly decoded bits, where 0.5 indicates a random output:
\begin{equation}
    \text{BAR} = \frac{\text{Number of Correctly Decoded Bits}}{\text{Total Number of Bits}}
\end{equation}
From this, we derive the \textbf{Watermark Uncertainty Score (WUS)}, which measures closeness to random noise (a score of 1 is a perfect attack):
\begin{equation}
    \text{WUS} = 1 - 2 \times |\text{BAR} - 0.5|
\end{equation}
Finally, to account for bit imbalances, we use the \textbf{Information Destruction Score (IDS)}, derived from the Matthews Correlation Coefficient (MCC). An IDS of 1 signifies that the statistical correlation between the original and extracted watermarks is completely eliminated:
\begin{equation}
    \text{IDS} = 1 - \left| \frac{\text{TP} \times \text{TN} - \text{FP} \times \text{FN}}{\sqrt{(\text{TP}+\text{FP})(\text{TP}+\text{FN})(\text{TN}+\text{FP})(\text{TN}+\text{FN})}} \right|
\end{equation}
where TP, TN, FP, and FN are the counts of True Positives, True Negatives, False Positives, and False Negatives for the decoded bits, respectively.

\begin{table}[h!]
\centering
\caption{Visual fidelity of the attacked 3DGS models using the single-objective \textbf{GMEA (w/o F1)} on the 1D watermarked Blender dataset. As expected, removing the visual quality objective ($F_1$) leads to a significant degradation in fidelity, reflected in the lower SSIM/PSNR and higher RMSE values. The arrows indicate the desired direction for each metric.}
\label{tab:visual_fidelity_1d_1obj_blender}
\begin{tabular}{@{}lccc@{}}
\toprule
\textbf{Scene} & \textbf{SSIM} $\uparrow$ & \textbf{PSNR} $\uparrow$ & \textbf{RMSE} $\downarrow$ \\
\midrule
Chair          & 79.31 & 19.12 & 0.1100 \\
Drums          & 83.24 & 20.16 & 0.0990 \\
Ficus          & 88.52 & 22.59 & 0.0747 \\
Hotdog         & 85.50 & 22.33 & 0.0784 \\
Lego           & 78.73 & 21.12 & 0.0899 \\
Materials      & 84.61 & 22.26 & 0.0784 \\
Mic            & 90.53 & 25.19 & 0.0554 \\
Ship           & 74.12 & 21.47 & 0.0855 \\
\midrule
\textbf{Average} & \textbf{83.07} & \textbf{21.78} & \textbf{0.0839} \\
\bottomrule
\end{tabular}
\end{table}

\begin{table}[h!]
\centering
\caption{Visual fidelity of the attacked 3DGS models (2D watermark, LLFF dataset) compared to the original watermarked models. Our full GMEA method successfully preserves high visual quality on these complex scenes, as indicated by the excellent metrics. The arrows indicate the desired direction for each metric.}
\label{tab:visual_fidelity_2d_llff}
\begin{tabular}{@{}lccc@{}}
\toprule
\textbf{Scene} & \textbf{SSIM} $\uparrow$ & \textbf{PSNR} $\uparrow$ & \textbf{RMSE} $\downarrow$ \\
\midrule
Fern           & 97.42 & 36.60 & 0.0148 \\
Flower         & 98.81 & 40.64 & 0.0094 \\
Fortress       & 98.73 & 39.95 & 0.0110 \\
Horns          & 98.57 & 39.05 & 0.0115 \\
Leaves         & 98.31 & 35.03 & 0.0178 \\
Trex           & 96.47 & 32.00 & 0.0252 \\
Room           & 98.64 & 39.60 & 0.0108 \\
Orchids        & 98.80 & 39.24 & 0.0110 \\
\midrule
\textbf{Average} & \textbf{98.22} & \textbf{37.76} & \textbf{0.0140} \\
\bottomrule
\end{tabular}
\end{table}

\begin{table}[h!]
\centering
\caption{Visual fidelity of the attacked 3DGS models using the single-objective \textbf{GMEA (w/o F1)} on the 1D watermarked LLFF dataset. This ablation study demonstrates a substantial degradation in visual quality, with very low SSIM/PSNR scores and high RMSE, confirming the necessity of the visual fidelity objective ($F_1$). The arrows indicate the desired direction for each metric.}
\label{tab:visual_fidelity_1d_1obj_llff}
\begin{tabular}{@{}lccc@{}}
\toprule
\textbf{Scene} & \textbf{SSIM} $\uparrow$ & \textbf{PSNR} $\uparrow$ & \textbf{RMSE} $\downarrow$ \\
\midrule
Fern           & 50.01 & 19.12 & 0.1106 \\
Flower         & 48.70 & 19.31 & 0.1083 \\
Fortress       & 56.58 & 19.81 & 0.1024 \\
Horns          & 53.87 & 19.01 & 0.1123 \\
Leaves         & 49.33 & 15.92 & 0.1602 \\
Trex           & 43.71 & 16.72 & 0.1460 \\
Room           & 65.70 & 20.81 & 0.0911 \\
Orchids        & 54.79 & 18.85 & 0.1143 \\
\midrule
\textbf{Average} & \textbf{52.84} & \textbf{18.69} & \textbf{0.1182} \\
\bottomrule
\end{tabular}
\end{table}

\begin{figure}[h!]
    \centering
    \includegraphics[width=1\linewidth]{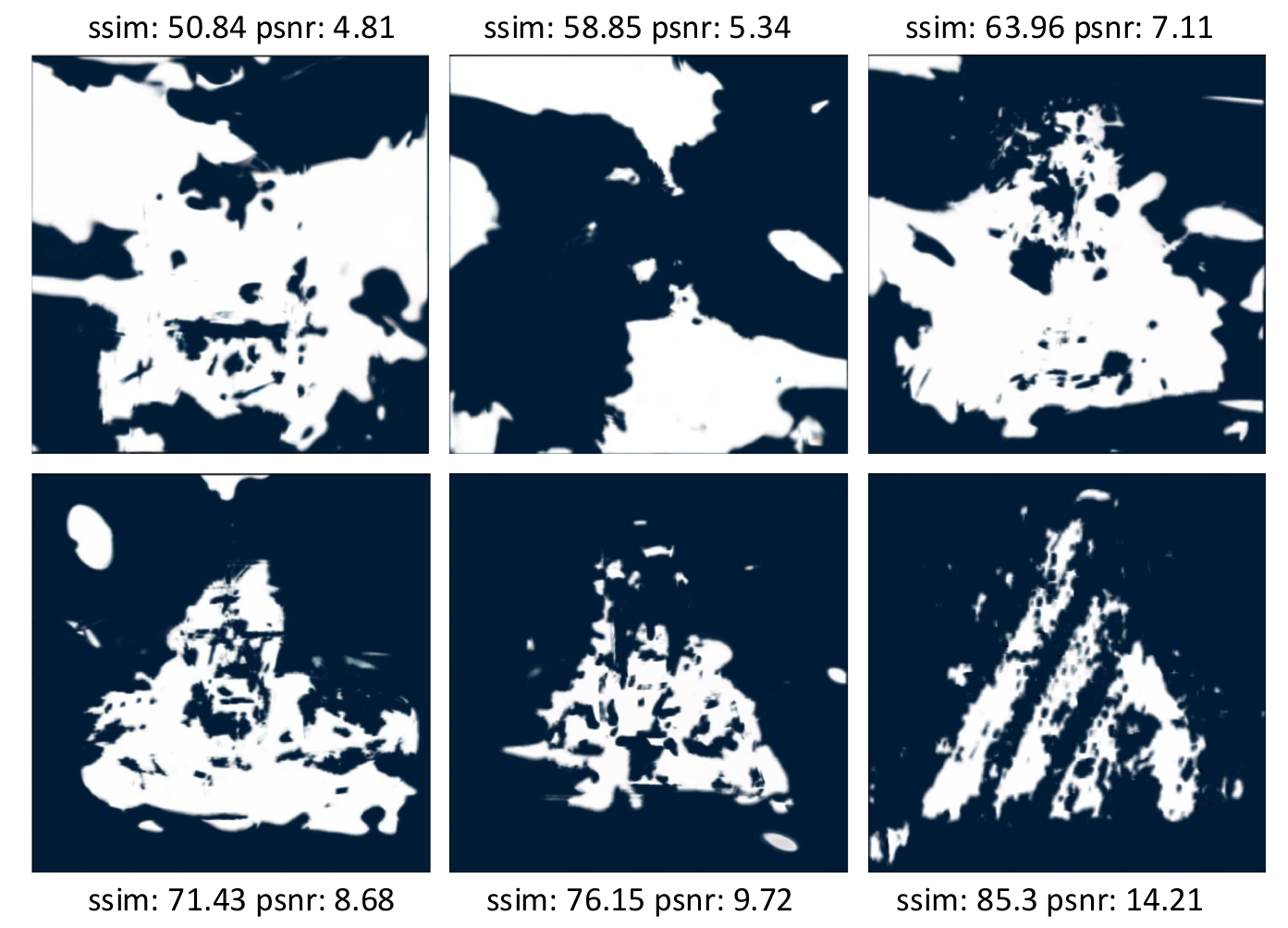}
    \caption{Qualitative results of 2D watermark extraction after the GMEA attack. The six examples illustrate the varying degrees of watermark corruption achieved by different solutions on the Pareto front, ranging from minimally distorted (corresponding to solutions that prioritize the 3DGS model's visual quality) to almost completely destroyed (corresponding to solutions that prioritize attack efficacy).}
    \label{fig:2d_watermark_corruption}
\end{figure}

\subsubsection{Group-Based Strategy Settings}
The partitioning of the 3DGS model was controlled by the following parameter:
\begin{itemize}
    \item \textbf{Number of Groups ($k$):} For our main experiments (Table 2 in main paper), we set the number of groups to \textbf{$k=10$}. This value was chosen based on our analysis (Table 3 in main paper), as it offers a significant reduction in computation time while keeping the peak memory usage manageable.
\end{itemize}

\subsubsection{Evolutionary Algorithm Settings}
Our attack is driven by a multi-objective evolutionary algorithm. Its parameters were set as follows:
\begin{itemize}
    \item \textbf{Population Size ($N_{pop}$):} We used a population of \textbf{$N_{pop}=50$} individuals for each sub-problem's evolutionary run.
    \item \textbf{Number of Generations ($T$):} The optimization was executed for a total of \textbf{$T=200$ generations}, which we observed was sufficient for the objective values to converge.
    \item \textbf{Color Perturbation ($\epsilon$):} The range for the color perturbation vector $\mathbf{c}^{(i)}$ was bounded by $[-\epsilon, \epsilon]$, with \textbf{$\epsilon = 50/255$}. 
    \item \textbf{Crossover Operator:} We used a simulated binary crossover (SBX) operator with a distribution index of \textbf{$\eta_c = 1.0$}.
    \item \textbf{Mutation Operator:} A polynomial mutation operator was applied with a mutation probability of \textbf{$p_m = 0.1$} and a distribution index of \textbf{$\eta_m = 20$}.
\end{itemize}

\subsubsection{Objective Function Settings}
The two competing objectives were configured with these parameters:
\begin{itemize}
    \item \textbf{Visual Fidelity ($F_1$):} The weight $\lambda$ in Equation 7 (main paper), which balances the L1 and SSIM losses, was set to \textbf{$\lambda = 0.85$}. The fitness evaluation was performed over a batch of \textbf{$N_v=8$} randomly sampled camera views in each generation.
    \item \textbf{Watermark Destruction ($F_2$):} The feature extractor $\Phi$ (Equation 9 in main paper) was a pre-trained \textbf{VGG-19 network}. We extracted features from the output of the \textbf{`relu4\_1` layer}, as its mid-level representational power is well-suited for capturing the patterns that constitute a watermark.
\end{itemize}

\subsubsection{Hardware and Software Environment}
All experiments were conducted on a high-performance computing server with the following specifications:
\begin{itemize}
    \item \textbf{CPU:} Intel(R) Xeon(R) Gold 5222 CPU @ 3.80GHz
    \item \textbf{GPU:} 2x NVIDIA A40 (48 GB VRAM each)
    \item \textbf{Operating System:} A Linux-based distribution
    \item \textbf{Core Libraries:} PyTorch was used for deep learning operations, including model rendering and feature extraction. The multi-objective evolutionary optimization was implemented using a standard Python library for evolutionary computation.
\end{itemize}

\subsection{Explanation of Algorithm 1: GMEA Framework}

To provide a clear and formal description of our attack framework, we present the corresponding pseudocode below. Our methodology consists of two main algorithms. Algorithm 1 outlines the high-level strategy of the Group-based Multi-objective Evolutionary Attack (GMEA), which partitions the problem and orchestrates the overall attack. Algorithm 2 details the core optimization process, an evolutionary attack applied to each sub-model, which is called within Algorithm 1. All the equations involved are those in the main paper.

Algorithm 1 describes the main workflow of our proposed GMEA.
\begin{itemize}
    \item \textbf{Step 1 (Line 3): Partitioning Phase.} The process begins by partitioning the entire 3DGS model, which can contain millions of Gaussians, into $k$ smaller, computationally manageable sub-models using K-Means. This step is crucial for making the large-scale optimization problem tractable.
    
    \item \textbf{Step 2 (Lines 4-6): Divide-and-Conquer.} The core of our framework is a divide-and-conquer strategy. We iterate through each sub-model and apply an independent optimization process, \texttt{EvolveSubModel} (detailed in Algorithm 2), to it. This design allows for massive parallelization, significantly speeding up the attack.
    
    \item \textbf{Step 3 (Line 7): Reconstruction.} Finally, after each sub-model has been optimized, they are merged back together to reconstruct the final adversarial 3DGS model, $\mathbf{G}_{adv}$.
\end{itemize}

\begin{algorithm}[H]
\caption{Group-based Multi-objective Evolutionary Attack (GMEA)}
\label{alg:gmea}
\begin{algorithmic}[1]
\State \textbf{Input:} Watermarked 3DGS model $\mathbf{G}_{wm}$, number of clusters $k$.
\State \textbf{Output:} Adversarial 3DGS model $\mathbf{G}_{adv}$.

\State Partition $\mathbf{G}_{wm}$ into $k$ sub-models $\{\mathbf{G}^{(1)}_{wm}, \dots, \mathbf{G}^{(k)}_{wm}\}$ using K-Means clustering with Eq. (2), (3), and (4).

\ForAll{sub-model $\mathbf{G}^{(i)}_{wm}$ in $\{\mathbf{G}^{(1)}_{wm}, \dots, \mathbf{G}^{(k)}_{wm}\}$}
    \State $\mathbf{G}^{(i)}_{adv} \leftarrow \text{EvolveSubModel}(\mathbf{G}^{(i)}_{wm})$ 
\EndFor

\State Reconstruct the final model $\mathbf{G}_{adv}$ by merging all optimized sub-models using Eq. (13).

\State \textbf{return} $\mathbf{G}_{adv}$.
\end{algorithmic}
\end{algorithm}

\begin{figure*}
    \centering
    \includegraphics[width=1\linewidth]{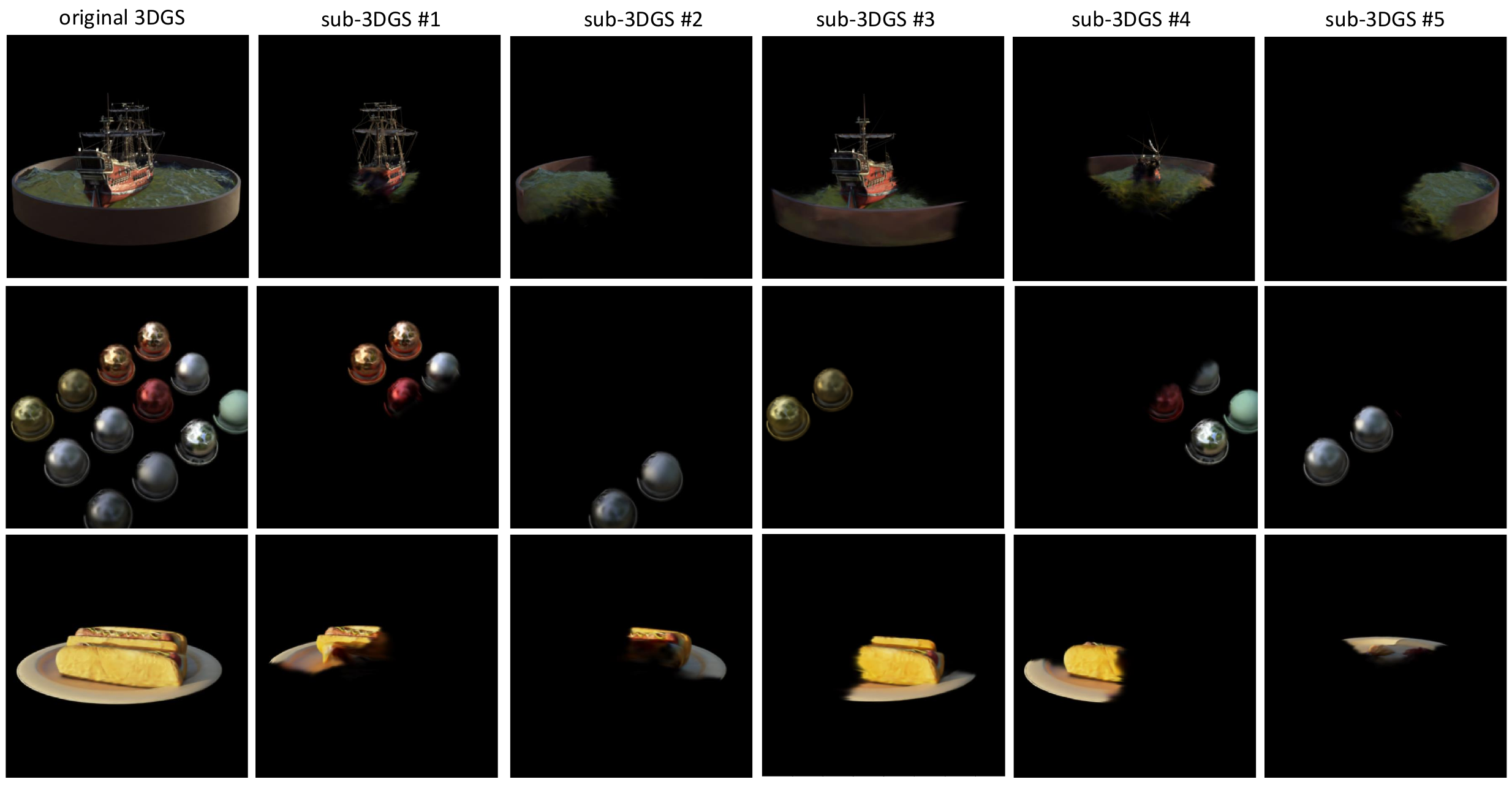}
    \caption{Visualization of our group-based partitioning strategy with $k=5$ on scenes from the \textbf{Blender dataset}. The leftmost column shows the render of the original, complete 3DGS model. The subsequent five columns display the individual sub-3DGS models, each representing a distinct and spatially coherent sub-problem for optimization.}
    \label{fig:group_split_blender}
\end{figure*}

\subsection{Explanation of Algorithm 2: Evolutionary Attack on a Sub-Model}

Algorithm 2 provides the implementation details for the \texttt{EvolveSubModel} function, which is the heart of our attack mechanism. Its goal is to find the optimal perturbation for a given sub-model to erase the watermark while preserving visual fidelity. All the equations involved are those in the main paper.
\begin{itemize}
    \item \textbf{Step 1 (Line 3): Population Initialization.} We initialize a population $\mathcal{P}_t$, where each individual $\mathbf{x}$ represents a potential set of modifications to the attributes (e.g., color, opacity, scale) of the Gaussians in the sub-model.
    
    \item \textbf{Step 2 (Lines 4-12): Evolutionary Loop.} The algorithm enters the main evolutionary loop. In each generation, we create new candidate solutions (offspring) using \textbf{crossover} and \textbf{mutation} operators.
    
    \item \textbf{Step 3 (Lines 7-10): Fitness Calculation.} Each candidate solution is evaluated using a fitness function with two competing objectives: $F_1$, which measures the success of watermark removal, and $F_2$, which quantifies the visual similarity to the original model.
    
    \item \textbf{Step 4 (Line 11): Environmental Selection.} A selection mechanism based on \textbf{non-dominated sorting} and a density metric (as used in NSGA-II) is employed to choose the individuals that will form the next generation's population. This preserves a diverse set of high-quality solutions that balance the two objectives.

    \item \textbf{Step 5 (Lines 13-14): Final Solution Selection.} After the final generation, the best solution $\mathbf{x}^*$ from the Pareto front is selected to construct the final optimized sub-model $\mathbf{G}^{(i)}_{adv}$.
\end{itemize}

\begin{algorithm}[H]
\caption{Evolutionary Attack on a Sub-Model}
\label{alg:sub_attack}
\begin{algorithmic}[1]
    \State \textbf{Input:} A watermarked 3DGS sub-model $\mathbf{G}^{(i)}_{wm}$.
    \State \textbf{Output:} An optimized adversarial sub-model $\mathbf{G}^{(i)}_{adv}$.

    \State Initialize population $\mathcal{P}_t$ with individuals $\mathbf{x}$ defined by Eq. (5).
    
    \For{$t = 1$ to $T$
    }
        \State Generate offspring $\mathcal{Q}_t$ from $\mathcal{P}_t$ using crossover (Eq. 10) and mutation (Eq. 11).
        \State Combine populations into a  pool $\mathcal{R}_t = \mathcal{P}_t \cup \mathcal{Q}_t$.
        
        \ForAll{individual $\mathbf{x}$ in $\mathcal{R}_t$}
            \State Construct candidate $\mathbf{G}'$ from $\mathbf{x}$ via Eq. (6).
            \State Calculate fitness ($F_1$, $F_2$) using Eq. (7), (8), (9).
        \EndFor
        
        \State Select next generation $\mathcal{P}_{t+1}$ via non-dominated sorting and density metric (Eq. 12).
    \EndFor
    
    \State Select best solution $\mathbf{x}^*$ from the final population $\mathcal{P}_{t+1}$.
    \State Construct final $\mathbf{G}^{(i)}_{adv}$ using the best solution $\mathbf{x}^*$.
    
    \State \textbf{return} $\mathbf{G}^{(i)}_{adv}$.

\end{algorithmic}
\end{algorithm}

\begin{figure*}
    \centering
    \includegraphics[width=1\linewidth]{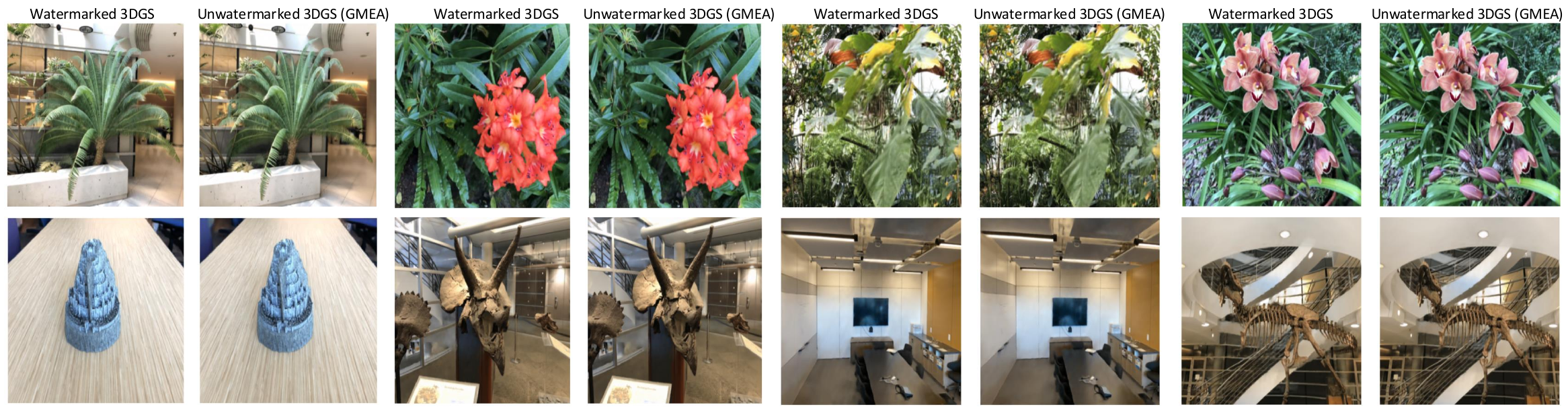}
    \caption{Qualitative comparison on the more complex \textbf{LLFF dataset}. Each pair consists of the original watermarked 3DGS render (left) and the unwatermarked version produced by our GMEA attack (right). Even in these challenging real-world scenes, the attacked renders are nearly indistinguishable from the originals, underscoring the effectiveness and robustness of our method in maintaining visual quality.}
    \label{fig:qualitative_llff}
\end{figure*}

\begin{table}[h!]
\centering
\caption{Visual fidelity of the attacked 3DGS models using the single-objective \textbf{GMEA (w/o F1)} on the 2D watermarked Blender dataset. The significant drop in SSIM/PSNR and the high RMSE values demonstrate the severe visual degradation when the fidelity-preserving objective ($F_1$) is omitted. The arrows indicate the desired direction for each metric.}
\label{tab:visual_fidelity_2d_1obj_blender}
\begin{tabular}{@{}lccc@{}}
\toprule
\textbf{Scene} & \textbf{SSIM} $\uparrow$ & \textbf{PSNR} $\uparrow$ & \textbf{RMSE} $\downarrow$ \\
\midrule
Chair          & 85.54 & 18.50 & 0.1207 \\
Drums          & 79.34 & 18.09 & 0.1251 \\
Ficus          & 86.09 & 19.64 & 0.1051 \\
Hotdog         & 86.50 & 21.90 & 0.0827 \\
Lego           & 72.87 & 17.20 & 0.1408 \\
Materials      & 81.30 & 20.40 & 0.0971 \\
Mic            & 86.27 & 21.18 & 0.2811 \\
Ship           & 70.93 & 20.57 & 0.0945 \\
\midrule
\textbf{Average} & \textbf{81.11} & \textbf{19.69} & \textbf{0.1309} \\
\bottomrule
\end{tabular}
\end{table}

\begin{figure}[h!]
    \centering
    \includegraphics[width=1\linewidth]{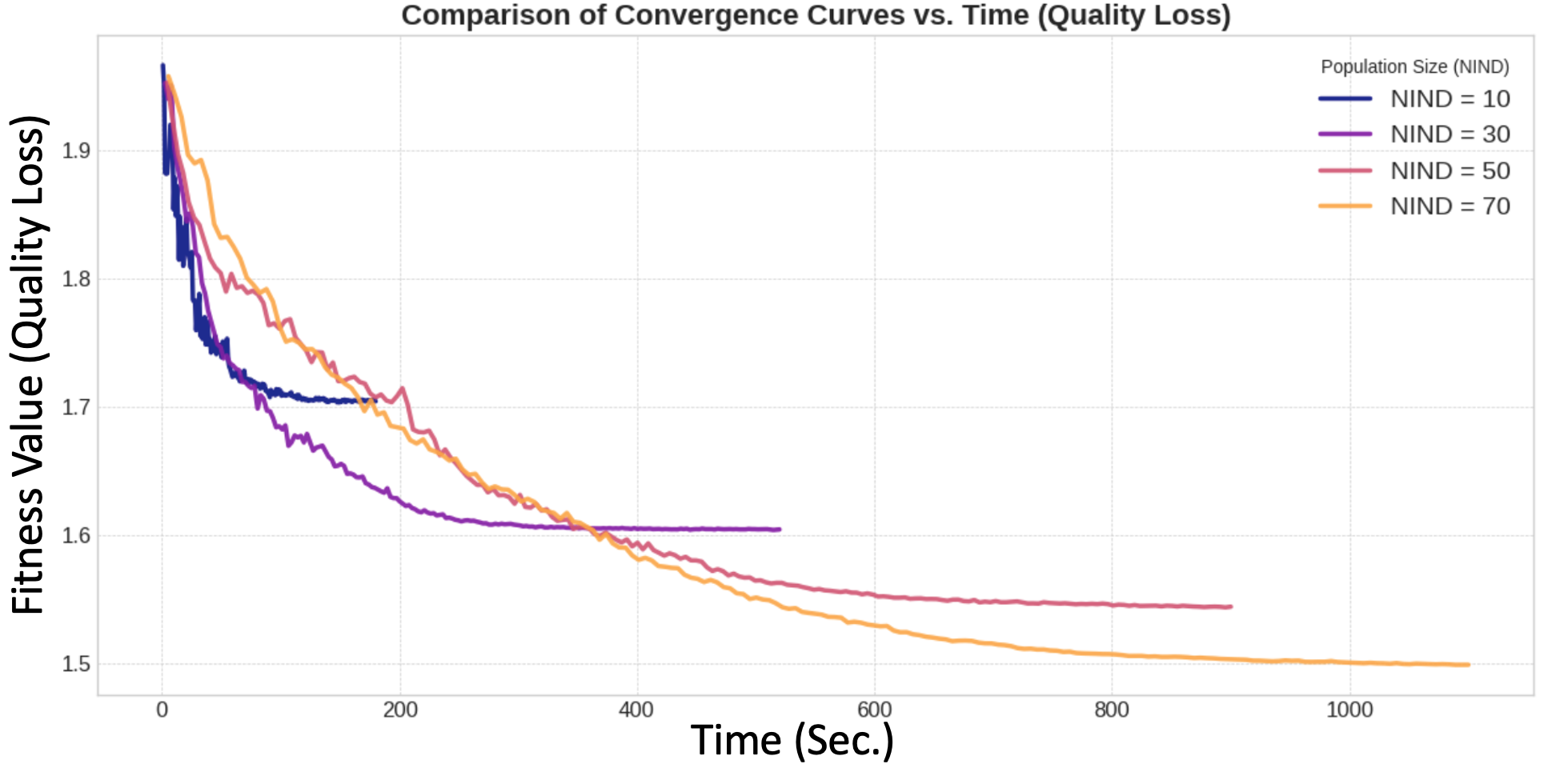}
    \caption{Convergence of Objective 1 (Quality Loss) vs. \textbf{Time (Sec.)}. This plot provides a fairer efficiency comparison. It reveals that while larger populations eventually find better solutions, a smaller population (e.g., NIND=30 or 50) can reach a good-quality solution much faster.}
    \label{fig:convergence_f1_vs_time}
\end{figure}

\section{Theoretical Justification}
To provide a theoretical foundation for our watermark destruction objective $F_2$, we introduce and prove Lemma 1. This lemma establishes a direct relationship between the variance of feature map activations and their information content. We model the activation values within a feature channel as a continuous random variable $Z$.

\newtheorem{lemma}{Lemma}
\begin{lemma}
For a continuous random variable with a given mean and variance, its differential entropy is upper-bounded by the entropy of a Gaussian distribution with the same mean and variance. This upper bound is a strictly monotonically increasing function of the variance. Consequently, minimizing the variance of the random variable forces a compression of its information entropy's upper bound, causing its distribution to degenerate into an uninformative Dirac delta function in the limit.
\end{lemma}

We divide this proof into four parts.

\subsubsection{A. The Gaussian Distribution Maximizes Entropy.}
We first prove that among all continuous probability distributions with a given mean $\mu$ and variance $\sigma^2$, the Gaussian distribution $\mathcal{N}(\mu, \sigma^2)$ has the maximum differential entropy.

Let $p(z)$ be an arbitrary probability density function (PDF) with mean $\mu$ and variance $\sigma^2$, and let $g(z)$ be the PDF of a Gaussian distribution with the same parameters. We consider the Kullback-Leibler (KL) divergence between these two distributions:
\begin{equation}
D_{KL}(p || g) = \int_{-\infty}^{\infty} p(z) \log \frac{p(z)}{g(z)} dz
\end{equation}
By Gibbs' inequality, we know that $D_{KL}(p || g) \ge 0$, with equality holding if and only if $p(z) = g(z)$. Expanding the definition of KL divergence:
\begin{equation}
D_{KL}(p || g) = \int_{-\infty}^{\infty} p(z) \log p(z) dz - \int_{-\infty}^{\infty} p(z) \log g(z) dz
\end{equation}
The first term is the negative differential entropy of $p(z)$, i.e., $-H(p)$. For the second term, the logarithm of the Gaussian PDF $g(z)$ is:
\begin{equation}
\begin{split}
    \log g(z) &= \log \left( \frac{1}{\sqrt{2\pi\sigma^2}} e^{-\frac{(z-\mu)^2}{2\sigma^2}} \right) \\
              &= -\frac{(z-\mu)^2}{2\sigma^2} - \log(\sqrt{2\pi\sigma^2})
\end{split}
\end{equation}
Substituting this into the integral of the second term:

\begin{align}
    &\int_{-\infty}^{\infty} p(z) \log g(z) dz \\
    &\quad= \int_{-\infty}^{\infty} p(z) \left( -\frac{(z-\mu)^2}{2\sigma^2} - \log(\sqrt{2\pi\sigma^2}) \right) dz \\
    &\quad= -\frac{1}{2\sigma^2} \int_{-\infty}^{\infty} p(z)(z-\mu)^2 dz \\
    &\qquad - \log(\sqrt{2\pi\sigma^2}) \int_{-\infty}^{\infty} p(z) dz
\end{align}

By definition, $\int p(z)(z-\mu)^2 dz = \sigma^2$ (the variance of $p(z)$) and $\int p(z) dz = 1$. This simplifies the expression to:
\begin{align}
    &-\frac{1}{2\sigma^2}(\sigma^2) - \log(\sqrt{2\pi\sigma^2}) \\
    &\qquad= -\frac{1}{2} - \frac{1}{2}\log(2\pi\sigma^2) \\
    &\qquad= -\frac{1}{2}\log(2\pi e \sigma^2)
\end{align}
This is precisely the negative differential entropy of the Gaussian distribution, $-H(g)$. The KL divergence thus becomes:
\begin{equation}
D_{KL}(p || g) = -H(p) - (-H(g)) = H(g) - H(p)
\end{equation}
Since $D_{KL}(p || g) \ge 0$, it follows that $H(g) - H(p) \ge 0$, which implies $H(p) \le H(g)$. This proves that the entropy of a Gaussian distribution is the upper bound for any distribution with the same variance. Thus, $H_{\text{max}}(Z) = H(g)$. 

\subsubsection{B. Monotonicity of the Entropy Bound w.r.t. Variance.}
Having established the entropy upper bound as $H_{\text{max}}(Z) = \frac{1}{2}\log(2\pi e \sigma^2)$, we now show its monotonicity with respect to the variance $\sigma^2$. We find the derivative of $H_{\text{max}}(Z)$ with respect to $\sigma^2$:
\begin{align}
    \frac{d H_{\text{max}}(Z)}{d (\sigma^2)} &= \frac{d}{d(\sigma^2)} \left( \frac{1}{2}\log(2\pi e \sigma^2) \right) \\
    &= \frac{1}{2} \cdot \frac{1}{2\pi e \sigma^2} \cdot (2\pi e) \\
    &= \frac{1}{2\sigma^2}
\end{align}
Since the variance $\sigma^2$ is strictly positive for any non-degenerate distribution ($\sigma^2 > 0$), the derivative $\frac{1}{2\sigma^2}$ is always positive. Therefore, $H_{\text{max}}(Z)$ is a strictly monotonically increasing function of the variance $\sigma^2$. 

\subsubsection{C. Degeneracy to the Dirac Delta Function in the Limit.}
Next, we analyze the behavior of the Gaussian distribution $\mathcal{N}(\mu, \sigma^2)$ as its variance approaches zero, i.e., $\sigma^2 \to 0$. The PDF is $g(z; \mu, \sigma) = \frac{1}{\sqrt{2\pi}\sigma} e^{-\frac{(z-\mu)^2}{2\sigma^2}}$. We examine its properties in the limit $\sigma \to 0^+$:
\begin{enumerate}
    \item \textbf{For $z \neq \mu$:} The term $(z-\mu)^2 > 0$. As $\sigma \to 0^+$, the exponent $-\frac{(z-\mu)^2}{2\sigma^2} \to -\infty$. Consequently, $\lim_{\sigma \to 0^+} g(z; \mu, \sigma) = 0$ for all $z \neq \mu$.
    \item \textbf{For $z = \mu$:} The exponent is $0$, making $e^0=1$. However, the leading coefficient $\frac{1}{\sqrt{2\pi}\sigma} \to +\infty$ as $\sigma \to 0^+$. Thus, $\lim_{\sigma \to 0^+} g(\mu; \mu, \sigma) = \infty$.
    \item \textbf{Integral Property:} For any $\sigma > 0$, the total integral of the PDF is unity: $\int_{-\infty}^{\infty} g(z; \mu, \sigma) dz = 1$. This property holds even in the limit.
\end{enumerate}
Collectively, these three properties—being zero everywhere except at a single point where it is infinite, while maintaining a total integral of one—are the defining characteristics of the \textbf{Dirac delta function}, $\delta(z-\mu)$. Thus, we formally deduce that $\lim_{\sigma \to 0^+} \mathcal{N}(\mu, \sigma^2) = \delta(z-\mu)$.

\begin{table}[h!]
\centering
\caption{Visual fidelity of the attacked 3DGS models using the single-objective \textbf{GMEA (w/o F1)} on the 2D watermarked LLFF dataset. The extremely low SSIM/PSNR scores and high RMSE values show that this attack variant severely compromises the visual integrity of the models, making the output unusable. The arrows indicate the desired direction for each metric.}
\label{tab:visual_fidelity_2d_1obj_llff}
\begin{tabular}{@{}lccc@{}}
\toprule
\textbf{Scene} & \textbf{SSIM} $\uparrow$ & \textbf{PSNR} $\uparrow$ & \textbf{RMSE} $\downarrow$ \\
\midrule
Fern           & 47.56 & 17.13 & 0.1392 \\
Flower         & 45.83 & 16.42 & 0.1512 \\
Fortress       & 29.49 & 12.19 & 0.2459 \\
Horns          & 53.55 & 16.89 & 0.1437 \\
Leaves         & 29.49 & 12.23 & 0.2447 \\
Trex           & 11.45 & 11.32 & 0.2718 \\
Room           & 66.26 & 18.54 & 0.1186 \\
Orchids        & 35.66 & 12.92 & 0.2260 \\
\midrule
\textbf{Average} & \textbf{39.91} & \textbf{14.71} & \textbf{0.1926} \\
\bottomrule
\end{tabular}
\end{table}

\subsubsection{D. Entropy in the Limit Case.}
Finally, we compute the limit of the entropy upper bound $H_{\text{max}}(Z)$ as $\sigma \to 0^+$:
\begin{equation}
\lim_{\sigma \to 0^+} H_{\text{max}}(Z) = \lim_{\sigma \to 0^+} \frac{1}{2}\log(2\pi e \sigma^2)
\end{equation}
As $\sigma \to 0^+$, the argument of the logarithm $2\pi e \sigma^2 \to 0^+$. Since $\lim_{x \to 0^+} \log(x) = -\infty$, it follows that:
\begin{equation}
\lim_{\sigma \to 0^+} H_{\text{max}}(Z) = -\infty
\end{equation}
This result indicates that as the variance approaches zero, the information entropy of the distribution approaches negative infinity. This signifies a complete absence of uncertainty—the system's state becomes perfectly deterministic and thus contains no information.

This proof rigorously establishes that minimizing the variance of feature map activations, as proposed in our objective function $F_2$, directly compresses the upper bound of their information entropy. This process forces the feature representation to become uniform and predictable, thereby destroying any complex patterns, including the embedded watermark, that rely on informational diversity.

\begin{figure}[h!]
    \centering
    \includegraphics[width=1\linewidth]{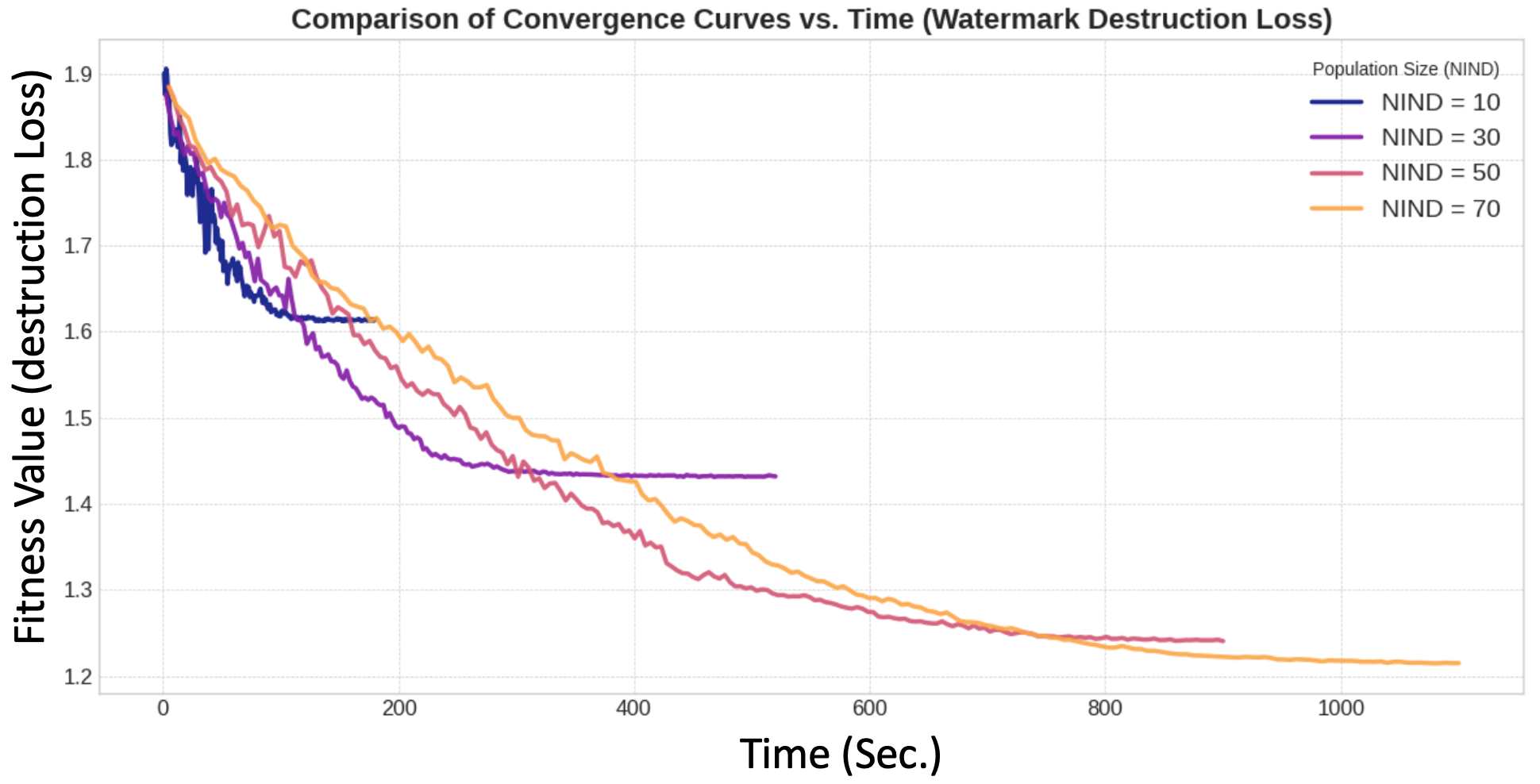}
    \caption{Convergence of Objective 2 (Watermark Destruction) vs. \textbf{Time (Sec.)}. This crucial comparison shows the time-based efficiency of watermark removal. It highlights that an intermediate population size offers the best trade-off, achieving significant watermark destruction in the shortest amount of time.}
    \label{fig:convergence_f2_vs_time}
\end{figure}

\section{Supplementary Experiments}

To complement the quantitative results in the main paper, we provide additional visual evidence and ablation studies that further validate the effectiveness and design choices of our GMEA framework.

\subsection{Qualitative Visual Results of GMEA }
Visual quality is paramount for a successful attack, as the goal is to obtain a high-fidelity, unwatermarked asset. To this end, we present qualitative comparisons on both the Blender and LLFF datasets to demonstrate the imperceptibility of our full GMEA framework.

\Cref{fig:qualitative_blender} showcases the results on various object-centric scenes from the \textbf{Blender dataset}. Each image pair compares the original watermarked render with the render after our attack. The visual differences are minimal, confirming that our method successfully preserves the model's visual integrity while removing the watermark.

To demonstrate the robustness of our approach on more challenging data, \Cref{fig:qualitative_llff} presents the same comparison on complex, real-world scenes from the \textbf{LLFF dataset}. Even with intricate lighting and geometry, our GMEA framework maintains exceptional visual fidelity, rendering the attack practically invisible to the human eye. This visual evidence strongly supports the quantitative metrics presented in the main paper and in our ablation tables, underscoring the stealth and effectiveness of our method.

\begin{table}[h!]
\centering
\caption{Computational cost analysis for different population sizes. The table shows the average wall-clock time required to compute a single generation for each population size (NIND) setting used in our experiments.}
\label{tab:time_per_generation}
\begin{tabular}{@{}cc@{}}
\toprule
\textbf{Population Size ($N_{pop}$)} & \textbf{Time per Generation (s)} \\
\midrule
70 & 5.5 \\
50 & 4.5 \\
30 & 2.6 \\
10 & 0.9 \\
\bottomrule
\end{tabular}
\end{table}

\subsection{Quantitative Analysis of Visual Fidelity}
In addition to the visual comparisons, we provide a detailed quantitative analysis of the visual fidelity of the attacked models across different settings. This analysis, presented in the tables below, serves two purposes: first, to numerically confirm the high quality preserved by our full GMEA framework, and second, to provide a thorough ablation study justifying our multi-objective design.

We begin by evaluating our complete two-objective GMEA framework. 
\Cref{tab:visual_fidelity_1d_blender} details the excellent visual quality metrics achieved when attacking 1D watermarked models on the Blender dataset. 
Similarly, \Cref{tab:visual_fidelity_2d_llff} shows the results for attacking 2D watermarked models on the more complex LLFF dataset. In both tables, the high average SSIM (above 95\%) and PSNR values, combined with low RMSE, provide strong quantitative evidence that our method preserves the asset's original visual quality.

To validate our choice of a multi-objective approach, we conducted an ablation study by removing the visual fidelity objective ($F_1$) and running a single-objective attack (GMEA (w/o F1)). The results of this study are detailed in the subsequent four tables.
\Cref{tab:visual_fidelity_1d_1obj_blender} and \Cref{tab:visual_fidelity_1d_1obj_llff} quantify the significant visual degradation when attacking 1D watermarks on the Blender and LLFF datasets, respectively. 
Likewise, \Cref{tab:visual_fidelity_2d_1obj_blender} and \Cref{tab:visual_fidelity_2d_1obj_llff} demonstrate a severe drop in quality for attacks on 2D watermarked models. 
Collectively, the low SSIM/PSNR scores and high RMSE values in these four tables confirm that a single-objective attack, while potent, renders the 3DGS asset visually unusable. This ablation study robustly justifies the necessity of our multi-objective formulation for any practical attack scenario where the goal is to steal a visually intact digital asset.

\subsection{Qualitative Analysis of 2D Watermark Corruption}
In addition to evaluating the visual quality of the attacked 3DGS models, we also provide a qualitative analysis of the 2D watermark's degradation. \Cref{fig:2d_watermark_corruption} showcases six examples of the extracted AAAI logo after our GMEA framework has been applied. These examples correspond to different optimal solutions found during the multi-objective optimization, each representing a unique trade-off between preserving the 3DGS model's visual fidelity ($F_1$) and maximizing watermark destruction ($F_2$). The results visually confirm our attack's effectiveness, demonstrating a spectrum of corruption from partially recognizable to almost completely erased. This supports the quantitative findings in the main paper, where lower SSIM/PSNR values for the extracted watermark indicate successful attacks.

\subsection{Visualization of the Group-Based Strategy}
\begin{figure*}
    \centering
    \includegraphics[width=1\linewidth]{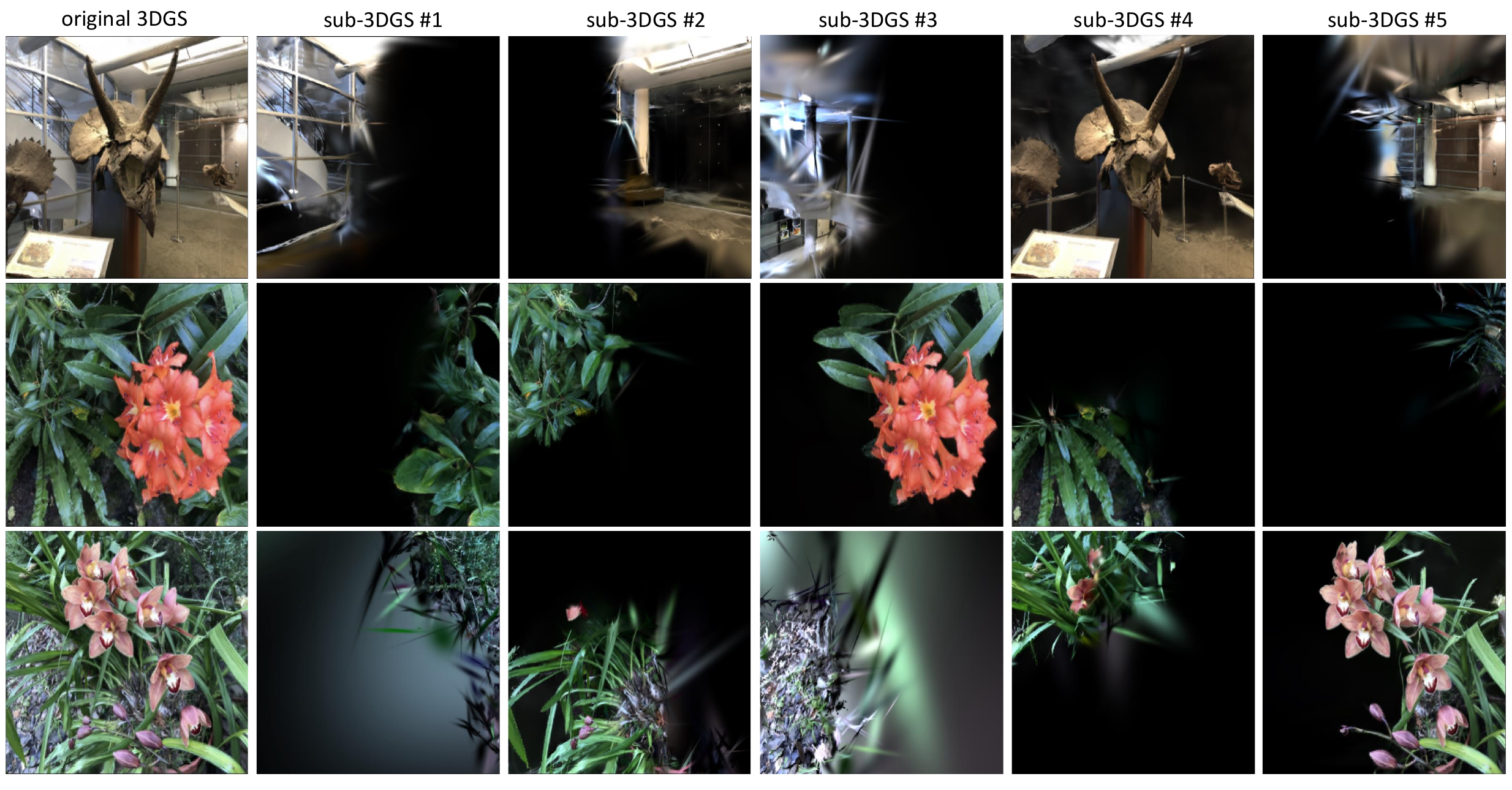}
    \caption{Visualization of our group-based partitioning strategy with $k=5$ on scenes from the more complex \textbf{LLFF dataset}. The first column shows the complete 3DGS model, while the following five columns show the spatially distinct sub-3DGS models produced by our clustering approach. This demonstrates the strategy's effectiveness in segmenting complex real-world scenes.}
    \label{fig:group_split_llff}
\end{figure*}

To provide an intuitive understanding of our group-based optimization strategy, we present visualizations of the partitioning process. \Cref{fig:group_split_blender} and \Cref{fig:group_split_llff} display the results of applying our strategy with $k=5$ to scenes from the Blender and LLFF datasets, respectively. In each figure, the first column shows the fully rendered original 3DGS model. The following five columns show the individual renders of each of the five sub-3DGS models generated by our K-Means clustering algorithm. As can be seen, the strategy effectively decomposes the complex scene into spatially coherent and localized sub-problems (e.g., separating the ship's sail from its hull, or isolating a specific cluster of flowers). This partitioning is the key to making the large-scale search space manageable and improving the efficiency of the subsequent multi-objective optimization.

\subsection{Ablation Study on Population Size}
To analyze the impact of the evolutionary algorithm's population size ($N_{pop}$ or NIND) on performance, we conducted an ablation study comparing four different settings: 10, 30, 50, and 70. A key consideration is the trade-off between the quality of the solution found per generation and the wall-clock time required to achieve it. We therefore present the convergence analysis from two distinct viewpoints.

First, we analyze convergence against the number of evolutionary steps. \Cref{fig:convergence_f1_vs_gen} and \Cref{fig:convergence_f2_vs_gen} show the performance of the two objective functions plotted against the \textbf{generation number}. These figures illustrate the raw evolutionary progress, where larger populations tend to explore the search space more effectively and achieve better fitness values at later generations.

Second, for a more practical assessment of efficiency, we analyze convergence against real-world time. This perspective is crucial because larger populations require significantly more computation per generation. We quantify this computational cost in \Cref{tab:time_per_generation}, which lists the average wall-clock time required to complete a single generation for each population size. Using this timing data, we re-plotted the convergence curves against \textbf{time in seconds}, as shown in \Cref{fig:convergence_f1_vs_time} and \Cref{fig:convergence_f2_vs_time}. These plots provide a fair comparison of how quickly each configuration can reach a satisfactory solution and help answer which population size is most efficient within a given time budget.

\section{Discussion}

Our work introduces GMEA, the first universal black-box attack framework for 3DGS watermarking, demonstrating significant vulnerabilities in current copyright protection schemes. The success of our multi-objective, group-based evolutionary approach underscores the need for more robust watermarking techniques. In this section, we discuss the current limitations of our method and outline promising directions for future research.

\subsection{Limitations}
While our group-based optimization strategy significantly accelerates the attack process, the framework still requires a moderate runtime. Depending on the model's complexity, a complete attack takes approximately 20 minutes to converge. This is due to the inherently iterative nature of the evolutionary algorithm, where each of the numerous fitness evaluations involves rendering the 3DGS model from multiple viewpoints and processing these images through a convolutional neural network.

However, it is crucial to contextualize this runtime. An attack on a digital asset is not a time-sensitive task, unlike applications such as real-time rendering. Given that the framework successfully removes the watermark in a challenging black-box setting while preserving high visual fidelity, we argue that this modest time cost is a highly acceptable trade-off. For a malicious actor, the ability to obtain a high-quality, unwatermarked asset makes this time investment a negligible factor.

\subsection{Future Work}
Building upon our findings, a key avenue for future research is the acceleration of the attack process to improve its efficiency. While not a critical limitation for the attack's purpose, reducing the computational overhead would make the framework more accessible and faster to deploy. We propose several promising directions:

\begin{itemize}
    \item \textbf{Surrogate-Assisted Optimization:} The most significant bottleneck is the fitness evaluation step. Future work could explore the use of surrogate models (or proxy models), such as small, lightweight neural networks. These models could be trained to approximate the expensive objective functions ($F_1$ and $F_2$) and would be used to pre-screen a large number of candidate solutions. The full, expensive evaluation would then be used only for the most promising individuals, drastically reducing the overall computational load.

    \item \textbf{Gradient Estimation in Black-Box Settings:} Although we cannot access the true gradients of the watermark detector, techniques for gradient estimation in black-box settings could be explored. Methods like Natural Evolution Strategies (NES) or finite-difference approximations could provide an estimated gradient to guide the search more directly, potentially leading to much faster convergence than the derivative-free approach of our current evolutionary algorithm.

    \item \textbf{Enhanced Parallelization:} Our group-based strategy already allows for a high degree of parallelization. This could be further enhanced by implementing a more sophisticated distributed computing framework, allowing the optimization of different sub-problems to scale across multiple machines in a network, rather than just multiple GPUs on a single server.
\end{itemize}

By pursuing these research directions, the efficiency of black-box attacks on 3DGS watermarking can be significantly improved, further highlighting the need for the development of more secure and robust copyright protection technologies.

\end{document}